\documentclass[useAMS,usenatbib,usegraphicx,referee]{mn2e}
\newcommand{\bugref}{\bibitem[\protect\citeauthoryear{dummy }{1893}]{dum}}

\title[Parsec-scale Circular Polarization of 41 AGN]
{The 15--43-GHz Parsec-scale Circular Polarization of 41 Active Galactic Nuclei}

\author[V. M. Vitrishchak et al.]{V. M. Vitrishchak$^{1}$, 
D. C. Gabuzda,$^{2}$
J. C. Algaba$^{2}$, E. A. Rastorgueva$^3$, 
\newauthor S. P. O'Sullivan$^{2}$ \& A. O'Dowd$^{2}$ \\
$^{1}$Sternberg Astronomical Institute, Moscow State
University, Universitetski{\u\i} prospekt 13, Moscow, 119992
Russia\\
$^{2}$Physics Department, University College Cork, Cork, Ireland\\
$^3$Tuorla Observatory, University of Turku, Turku, Finland}

\begin{document}

\date{}
\pagerange{\pageref{firstpage}--\pageref{lastpage}} \pubyear{2008}
\maketitle
\label{firstpage}
\begin{abstract}
We present the results of parsec-scale circular polarization measurements
based on Very Long Baseline Array data for a number of radio-bright,
core-dominated Active Galactic Nuclei obtained simultaneously at
15, 22 and 43 GHz. The degrees of circular polarization $m_c$ for the VLBI core
region at 15 GHz are similar to values reported earlier at this wavelength,
with typical values of a few tenths of a percent. 
We find that 
$m_c$ as often rises as falls with increasing frequency between 15 and 22~GHz,
while the degree of circular polarisation at 43~GHz is in all cases higher 
than at 22 and 15~GHz. This behaviour seems contrary to expectations,
since the degree of circular polarization from both synchrotron radiation
and Faraday conversion of linear to circular polarization -- the two main
mechanisms considered thus far in the literature -- should {\em decrease} 
towards higher frequencies if the source is homogeneous. The increase in 
$m_c$ at 43~GHz may be due to the presence of regions of both 
positive and negative circular polarisation with different frequency 
dependences (but decreasing with increasing frequency) on small scales 
within the core region;  alternatively, it may be associated with the
intrinsic inhomogeneity of a Blandford--K\"onigl-like jet. In several 
objects, the detected circular polarization appears to be near, but not 
coincident with the core, although further observations are needed to
confirm this.  We find several cases of changes in sign with frequency, 
most often between 22 and 43~GHz.  We find tentative evidence for 
transverse structure in the circular polarization of 1055+018 and 
1334$-$127 that is consistent with their being generated by either the
synchrotron mechanism or Faraday conversion in a helical magnetic field. 
Our results confirm the earlier finding that the sign of the
circular polarization at a given observing frequency is generally consistent
across epochs separated by several years or more, suggesting stability of
the magnetic field orientation in the innermost jets. 
\end{abstract}

\section{Introduction}

The radio emission of core-dominated, radio-loud Active Galactic Nuclei
(AGN) is synchrotron radiation
generated in the relativistic jets that emerge from the nucleus of the
galaxy, presumably along the rotational
axis of a central supermassive black hole. Synchrotron radiation can be
highly linearly polarized, to $\simeq 75\%$ in the case of a uniform magnetic
({\bf B}) field (Pacholczyk 1970), and linear polarization observations
can yield unique information about the orientation and degree of order of the
{\bf B} field in the synchrotron source, as well as the distribution of
thermal electrons and the {\bf B}-field geometry in the immediate vicinity
of the AGN (e.g., via Faraday rotation of the plane of polarization).

Techniques for deriving circular-polarization (CP) information on parsec scales
were pioneered by Homan and his collaborators in the late 1990s (Homan \& 
Wardle 1999; Homan, Attridge \& Wardle 2001) using data taken on the 
NRAO\footnote{The National Radio Astronomy Observatory of
the USA is operated by Associated Universities, Inc., under
co-operative agreement with the US NSF.} Very Long Baseline
Array (VLBA). Recently, CP measurements for the first
epoch of the MOJAVE project (monitoring of 133 AGN at 15 GHz with the VLBA)
were published by Homan and Lister (2006). Circular polarization was
detected in 34 of these objects at the 2$\sigma$ level or higher. These
results confirmed previously noted trends: the circular polarization is nearly
always coincident with the VLBI core, with typical degrees of polarization
$m_c$ being a few tenths of a percent.  Homan \& Lister (2006)
found no evidence for any correlation between $m_c$ and any of 20 different
optical, radio and intrinsic parameters of the AGN. Interestingly, five of
the 34 AGN displayed CP in their {\em jets}, well outside
the VLBI-core region, suggesting that the mechanism generating the circular
polarization is capable of operating effectively in optically thin regions
(although, strictly speaking, direct spectral-index measurements were not 
available at the corresponding epoch, since the MOJAVE measurements were made 
only at 15~GHz).

The two main mechanisms that are usually considered to be the most likely
generators of the observed CP are the synchrotron mechanism and the Faraday
conversion of linear to circular polarization [Legg \& Westfold 1968,
Jones \& O'Dell 1977; see also the reviews by Beckert \& Falcke 
(2002) and Wardle \& Homan (2003)].  Although the intrinsic CP generated by 
synchrotron radiation may be able to reach a few tenths of a percent at
15~GHz for the magnetic-field strengths characteristic of the observed
VLBI cores of AGN (typically $\simeq 0.4$~G; Lobanov 1998, O'Sullivan \& 
Gabuzda, in preparation), the highest observed $m_c$ values seem too high 
to plausibly be attributed to this mechanism. This suggests that Faraday 
conversion plays a role, possibly the dominant one, since it is expected to
be more efficient at generating CP than the synchrotron mechanism for 
the conditions in radio cores (Jones \& O'Dell 1977).  

Faraday conversion (Jones \& O'Dell 1977, Jones 1988) occurs because the
component of the linear-polarization electric vector parallel to the
conversion magnetic field, E$_{\|}$, gives rise to oscillations of free
charges in the conversion region, while the component orthogonal to this
magnetic field, E$_{\perp}$, cannot (the charges are not free to move
orthogonal to the magnetic field). This leads to a delay of E$_{\|}$
relative to E$_{\perp}$, manifest as the introduction of a small amount of
circular polarization. We recently obtained intriguing evidence that the 
CP in AGN may be generated by Faraday conversion in helical jet magnetic 
fields (Gabuzda et al. 2008). In this scenario, linearly
polarized radiation emitted at the far side of the jet is partially converted
to circular polarization as it passes through the magnetic field at the near
side of the jet on its way to the observer. In the simple model considered, 
the sign of the generated CP is determined by the pitch angle of the
helical field, which can be approximately deduced from the observed linear
polarization structure, and the helicity of the helical field, which can
be deduced from the direction of the Faraday-rotation gradient across the
VLBI jet, due to the systematically changing line-of-sight component of
the helical field. The inferred helicity depends on whether the longitudinal
component of the field corresponds to a ``North'' or ``South'' pole of the
central black hole.  After identifying 8 AGN with both detected VLBI-scale CP
and transverse Faraday rotation gradients, Gabuzda et al. (2008) determined 
the CP signs that should be observed for the cores of these AGN based on the
inferred pitch angles and the helicities for their helical fields, if the
jets were emerging from the North or South magnetic pole of the black hole.
The result was clearly not random: in all 8 cases, the observed CP signs 
agreed with the signs predicted by the simple helical-field model for the
case of the longitudinal field corresponding to South polarity. Although the
physical origin of this result is not clear, it strongly suggests
a close connection between CP and the presence of helical jet magnetic fields;
the probability of finding that 8 of 8 of these AGN had CP signs corresponding
to one particular polarity is less than 1\%. 
The apparent predominance of jets associated with South magnetic poles
may have cardinal implications for the intrinsic magnetic-field
structures of the central black holes; for example, one way to understand
this result is if the central fields are quadrupolar, with a predomint
tendency for the two South poles to correspond to the jets and the two
North poles to lie in the plane of the accretion disc, for
reasons that are not yet understood.

We present here new CP measurements for 41 AGN at 15, 22 and 43~GHz, as well
as for an additional 18 AGN at 15~GHz only. Parsec-scale CP was detected at
two of the three frequencies in 5 of the 41 AGN, and at all three frequencies 
in another 5 AGN.
Our 15-GHz results are generally in excellent agreement with the first-epoch
MOJAVE CP values (Homan \& Lister 2006).  The results do not show any 
universal frequency dependence for the degree of CP $m_c$, and both rising
and falling spectra with frequency are observed. Unexpectedly, we find
$m_c$ to be higher at 43~GHz than at the lower two frequencies. We also 
find tentative evidence for transverse CP structure consistent with the
CP being produced in a helical magnetic field geometry in two objects. 

\section{The Data and their Reduction}

We derived our multi-frequency CP measurements from simultaneous 
15.285, 22.285 and 43.131~GHz VLBA polarization observations of compact 
AGN obtained 
at epochs 7 August 2002, 5 March 2003, 1 November 2004, 15 March 2005 
and 26 September 2005. We also present 15-GHz results derived from data
obtained on 27 December 1999. The data were all obtained in a
``snapshot'' mode, with about 10 scans of 3--4~min duration of each source
at each frequency spread throughout the time the object was visible with 
all or nearly all of the VLBA antennas. The preliminary calibration,
D-term calibration and imaging were done in the NRAO AIPS package 
following standard techniques; for further information about the calibration 
procedures used, see Gabuzda et al. (2006), where some total intensity,
linear polarization and rotation-measure maps for objects observed
in August 2002 and March 2003 are presented.

The CP calibration of the gains was done using the well established 
gain-transfer (GT) technique of Homan and Wardle (1999), described in detail 
both in that paper and by Vitrishchak \& Gabuzda (2007).  In this method,
estimates of the right-to-left (R/L) gain ratios for various sources at various 
times are collated and smoothed to obtain a master R/L gain calibration 
table, which is then applied to the data for each source. Sources known or
suspected of having detectable CP are excluded from those used to make
this master calibration table. In all cases, we applied a Gaussian smoothing 
function with a full width at half maximum of four hours. The successful 
application of this
technique depends on observing a sufficiently large number of sources over a
sufficiently long time. This criterion was met reasonably well by all our
observations, as we typically observed 10--12 objects in 24 hours in each 
experiment.

The $1\sigma$ uncertainties in the CP values were determined in 
essentially the same way as is described by Homan, Attridge \& Wardle (2001),
including contributions for uncertainty (i) in the smoothed antenna gains,
(ii) due to real CP in the calibrators, and (iii) due to
random scan-to-scan gain variations. We also added the fractional
CP corresponding to the rms in the $V$ map in quadrature. 
This uncertainty-estimation procedure 
is relatively conservative, and we are in the course of investigating a
Monte Carlo approach to estimating the CP uncertainties, similar to that 
used by Homan \& Lister (2006).

\begin{table*}
\caption{CP Coincident with the VLBI Core} \centering
\label{tab:cp_core}
\begin{tabular}{lccccccccc}
\hline
Source   &  $z$ & Optical Class & Epoch & \multicolumn{2}{c}{15~GHz} & \multicolumn{2}{c}{22 GHz} & \multicolumn{2}{c}{43 GHz} \\
 &&&   & $m_c$ (\%)& $\sigma$ & $m_c$ (\%)& $\sigma$ & $m_c$ (\%)& $\sigma$\\
0048$-$097& $\cdots$ &  B & 05 Mar 2003 & $<0.28$   & $-$ & $<0.76$ & $-$&$<1.24$ & $-$\\ 
         &    &    & 01 Nov 2004 & $< 0.26$  & $-$ & $<0.74$ & $-$ & $<1.10$ & $-$ \\
0109+224 & $\cdots$  &  B & 27 Dec 1999 & $-0.24\pm0.09$ &2.7 & -- & -- & -- & -- \\
0133+476 &0.859& Q & 26 Sept 2005& $-0.32\pm0.09$ & 3.6 & $<0.28$ &$-$ & $-0.43\pm0.19$& 2.3\\
0138$-$097& 0.733 &  B & 05 Mar 2003 & $<0.42$   & $-$ & $<1.28$& $-$& $<7.42$& $-$\\
          &    &    & 26 Sept 2005& $<0.44$   & $-$ &$<1.40$ & $-$&$<1.30$ & $-$\\
0215+015 &1.715& B & 27 Dec 1999 & $<0.16$ &  $-$& -- & -- & -- & -- \\
0256+075 & 0.893 &  B & 05 Mar 2003 & $<0.52$  & $-$ & $<1.62$ & $-$ & -- & -- \\ 
         &    &    & 01 Nov 2004 & $< 0.50$ & $-$ & $< 1.36$ & $-$ & $<3.34$& $-$ \\ 
0300+470 &$\cdots$&B & 27 Dec 1999 & $-0.30\pm0.10$ &3.0 & -- & -- & -- & -- \\
0306+102 & 0.863 &  B & 27 Dec 1999 & $<0.18$ & $-$ & -- & -- & -- & -- \\
0420$-$014& 0.915 &  Q & 15 Mar 2005 & $<0.16$   & $-$ & $<0.30$ & $-$ & $<0.50$ & $-$ \\
0422+004 &$\cdots$& B & 27 Dec 1999 & $<0.24$ & $-$ & -- & -- & -- & -- \\
0735+178 &$\cdots$&  B & 05 Mar 2003 & $<0.22$ & $-$ & $<0.54$ & $-$& $<1.00$ & $-$\\
0745+241 & 0.409  &  B & 15 Mar 2005 & $<0.26$ & $-$ & $<0.50$ &$-$  & $<1.12$& $-$ \\
0754+100 &0.266& B & 27 Dec 1999 & $<0.20$ & $-$ & -- & -- & -- & -- \\
0804+499 &1.432& Q & 01 Nov 2004 &  $< 0.20$ & $-$ & $<0.54$ & $-$ & $<0.81$ & $-$ \\ 
0808+019 &$\cdots$& B & 27 Dec 1999 & $<0.26$ & $-$ & -- & -- & -- & -- \\
0814+425 &$\cdots$& B  & 26 Sept 2005& $<0.30$ & $-$ & $<0.80$& $-$&$<1.28$ & $-$\\
0818$-$128 & $\cdots$&  B & 27 Dec 1999 & $<0.66$ & $-$ & -- & -- & -- & -- \\
0823+033 &0.506&  B & 05 Mar 2003 & $+0.20\pm0.09$ & 2.2 & $<0.48$&$-$ & $<0.76$&$-$ \\
0829+046 & 0.174 &  B & 27 Dec 1999 & $<0.22$ & $-$ & -- & -- & -- & -- \\
0851+202 & 0.306 &  B & 05 Mar 2003 & $-0.15\pm0.08$ & 2.0 & $<0.40$& $-$ &$<0.48$ &  $-$\\
         &       &    & 26 Sept 2005& $-0.19\pm0.08$ & 2.3 & $-0.22\pm0.13$& 1.7&$+0.49\pm0.26$ & 1.9\\
0859+470 & 1.470 & Q  & 15 Mar 2005 & $<0.24$  & $-$ & $<0.44$ & $-$ & $<1.48$ & $-$ \\
0906+430 & 0.670 & Q  & 01 Nov 2004 & $< 0.18$ & $-$ & $<0.46$ &$-$ &$<0.84$ & $-$\\
0953+254 & 0.712 & Q  & 15 Mar 2005 & $<0.38$  & $-$ & $<0.58$ & $-$ & $<0.78$& $-$ \\
1055+018 & 0.888 &  Q & 15 Mar 2005 & $+0.52\pm 0.10$ & 5.0 & $+0.29\pm0.17$& 1.7 &See Table~\ref{tab:cp_noncore} & \\
1034$-$293 & 0.312 &  B & 27 Dec 1999 & $+0.37\pm0.12$ & 3.1 & -- & -- & -- & -- \\
1147+245 & 0.200&  B & 05 Mar 2003 & $<0.28$ & $-$ & $<0.78$& $-$& $<1.64$& $-$\\
1156+295 & 0.729&  Q & 07 Aug 2002 & $-0.28\pm0.11$ & 2.6 & $<0.58$& $-$ &$<1.28$ & $-$ \\
         &    &    & 01 Nov 2004 &  $< 0.24$ & $-$ &$<0.48$ & $-$ & $<0.82$& $-$ \\
         &    &    & 26 Sept 2005&  $<0.16$  & $-$ & $<0.42$& $-$& $<0.44$ & $-$\\
1215+303 & 0.130&  B & 27 Dec 1999 & $<0.36$ & $-$ & -- & -- & -- & -- \\
1253$-$055 & 0.538 &  Q & 07 Aug 2002 & $+0.19\pm0.11$ & 1.7 & $<0.80$&$-$ & $<1.44$& $-$\\
           &       &    & 04 Mar 2003 & $+0.83\pm 0.10$ & 8.3 &$+0.62\pm0.25$ &2.5 & $+1.21\pm0.37$& 3.3\\
           &       &    & 15 Mar 2005 & $+0.26\pm0.09$ & 2.9 & $+0.21\pm0.15$ & 1.4 & $-0.98\pm 0.16$ & 6.1\\
1307+121 &$\cdots$&  B & 27 Dec 1999 & $<0.24$ & $-$ & -- & -- & -- & -- \\
1334$-$127 & 0.539 &  Q & 05 Mar 2003 & $+0.28\pm0.09$ & 3.0 & $+0.44\pm0.24$& 1.8 & See Table~\ref{tab:cp_noncore} &\\
1413+135 &0.247&  B & 27 Dec 1999 & $<0.28$   & $-$ & -- & -- & -- & -- \\
1418+546 &0.153 &  B & 07 Aug 2002 & $<0.30$   & $-$& $<1.04$ & $-$& $<2.40$& $-$\\
1510$-$089 & 0.36 &  Q & 15 Mar 2005 & $<0.22$ &$-$ & $+0.49\pm0.19$& 2.6& $-2.84\pm0.40$ & 7.1 \\
1514+197 &1.070 &  B & 27 Dec 1999 & $<0.20$  & $-$ & -- & -- & -- & -- \\
1514$-$241 &0.049&  B & 27 Dec 1999 & $<0.22$ & $-$ & -- & -- & -- & -- \\
1538+149 &0.605&  B & 07 Aug 2002 & $<0.22$ & $-$ & $<0.72$ & $-$ & $<0.98$& $-$ \\
         &       &    & 05 Mar 2003 & $<0.22$ & $-$ &$<0.60$ & $-$&$<0.96$ & $-$\\
1611+343 &1.401&  Q & 26 Sept 2005& $<0.14$   & $-$ & $<0.24$ & $-$& $<0.40$ & $-$\\
1633+382 &1.807&  Q & 01 Nov 2004 &  $-0.34\pm 0.06$ & 5.7 & $-0.86\pm 0.17$&5.0 &$<0.54$ & $-$\\
1637+574 &0.751& Q & 15 Mar 2005 & $<0.24$ & $-$ & $<0.48$ & $-$ & $<0.80$ & $-$ \\
1641+399 &0.594 & Q & 01 Nov 2004 &  $<0.12$ & $-$ &$<0.38$ & $-$ & $<0.68$& $-$ \\
1652+398 & 0.034 &  B & 26 Sept 2005& $<0.32$ & $-$ & $<1.00$ & $-$&$<4.16$ & $-$\\
1717+178 &0.137 &  B & 27 Dec 1999 & $<0.26$ & $-$ & -- & -- & -- & -- \\
1732+389 &0.970 &  B & 07 Aug 2002 & $<0.26$ & $-$ & $<0.76$ &$-$ & $<0.94$& $-$\\
         &    &    & 05 Mar 2003 & $<0.26$ & $-$ & $<0.80$& $-$& $<1.02$& $-$\\
1739+522 &1.379& Q  & 15 Mar 2005 & $<0.22$ & $-$ & $<0.46$ & $-$ & $<1.68$& $-$ \\
1749+096 &0.322&  B & 27 Dec 1999 & $-0.19\pm0.08$ & 2.4 & -- & -- & -- & -- \\
         &    &    & 07 Aug 2002 & $-0.21\pm0.08$ & 2.5 & $<0.54$ & $-$& $<1.04$ & $-$\\
1823+568 &0.663&  B & 27 Dec 1999 & $<0.18$ & $-$ & -- & -- & -- & -- \\
         &    &    & 07 Aug 2002 & $<0.22$  & $-$ & $<0.78$ & $-$&$<1.18$ & $-$\\
1954+513 &1.220& Q & 15 Mar 2005 & $<0.32$ & $-$ & $<0.56$ & $-$ & $<1.50$& $-$ \\
\hline
\end{tabular}
\end{table*}

\begin{table*}
\addtocounter{table}{-1}
\caption{CP Coincident with the VLBI Core (cont'd)} \centering
\begin{tabular}{lccccccccc}
\hline
Source   &  $z$ & Optical Class & Epoch & \multicolumn{2}{c}{15~GHz} & \multicolumn{2}{c}{22 GHz} & \multicolumn{2}{c}{43 GHz} \\
 &&&   & $m_c$ (\%)& $\sigma$ & $m_c$ (\%)& $\sigma$ & $m_c$ (\%)& $\sigma$\\
2032+107 & 0.601&  B & 27 Dec 1999 & $<0.36$ & $-$ & -- & -- & -- & -- \\
2131$-$021 &1.285 &  B & 07 Aug 2002 & $<0.24$ & $-$ & $<0.70$ & $-$&  $<0.78$& $-$\\
2134+004 &1.932 &  Q & 01 Nov 2004 &  $< 0.22$ & $-$ &$<0.52$ & $-$ & $<1.12$& $-$ \\
2145+067 & 0.999&  Q & 26 Sept 2005& $-0.45\pm0.09$ & 5.0 & $-0.30\pm0.13$ & 2.3 & $<0.44$ & $-$ \\
2155$-$155 &0.672 &  Q & 26 Sept 2005& $<0.26$ & $-$ & $<0.68$ & $-$ & $<0.98$ & $-$ \\
2200+420 & 0.069 &  B & 07 Aug 2002 & $<0.16$ & $-$ & $<0.48$& $-$&$<0.96$ & $-$\\
2223$-$052 &1.404 &  Q & 27 Dec 1999 & $-0.20\pm0.07$ & 2.8 & -- & -- & -- & -- \\
2230+114 &1.037 &  Q & 01 Nov 2004 &  $-0.61\pm 0.08$ & 7.2 &$-1.38\pm0.21$ & 6.6 &$<0.52$ & $-$ \\
2251+158 & 0.859 & Q  & 26 Sept 2005& See Table~\ref{tab:cp_noncore}  &   & $-0.29\pm0.12$& 2.4 & $+0.29\pm0.18$& 1.6 \\
2254+074 & 0.190&  B & 07 Aug 2002 & $<0.36$& $-$ & $<1.22$&$-$ &$<1.72$ & $-$\\
\hline
\end{tabular}
\end{table*}

\section{Results}

Table~1 gives a list of the AGN observed in the 6 experiments, together
with their redshifts, optical classes, the epochs at which they were
observed, and our estimated degrees of core CP at
each of the three frequencies, $m_c$, obtained by taking the ratio of 
the peaks of the Stokes $V$ (circular polarization) and $I$ (total intensity)
images (when the $V$ and $I$ peaks were coincident or nearly coincident). 
These values for $m_c$ were derived using the beam sizes appropriate for 
each frequency.  The $1\sigma$ errors are also
indicated; when CP was not detected, $2\sigma$ is given as
the upper limit. A dash in the 22 and 43-GHz columns for $m_c$ and $\sigma$
indicates that the source was not observed at those frequencies 
for the indicated experiment. 

Table~1 lists several tentative measurements with significances between 
$1\sigma$ and $2\sigma$. We have included these when there was some 
supporting evidence of their reality, in particular, detection of CP at 
the other frequencies observed or detection of CP with the same sign in 
the first-epoch MOJAVE experiments (Homan \& Lister 2006). However, it 
should be borne in mind that these should not be considered firm detections 
until they are confirmed by further measurements, and the skeptical
reader should feel free to consider these upper limits equal to
the indicated CP values plus $1\sigma$.  All of the objects with 
measured CP
in Table~1 have CP measurements with significances of $2\sigma$ or higher 
at at least one frequency.  

The BL Lac object 1034$-$293 was the only object for which we detected 
CP that is not included in the MOJAVE sample.  Table~2 gives a comparison 
of our results for the remaining 16 sources with the 
first-epoch 15-GHz MOJAVE results (Homan \& Lister 2006). In all 8 cases
when CP is detected in both our experiments and the MOJAVE first-epoch
experiments, the sign of the CP at 15 GHz for the two datasets agree.
We did not detect 15-GHz CP in 1510$-$089, but the sign of CP we detect
at 22~GHz agrees with the sign of the MOJAVE measurements. In several
other sources, our detections are consistent with the MOJAVE upper limits.
This all shows the reliability of our measurements, and demonstrates that a
self-consistent picture is emerging from the accumulating data, despite
the difficulty of these measurements.  

\begin{table*}
\caption{Comparison with MOJAVE CP} \centering
\label{tab:mojave}
\begin{tabular}{lccl}
\hline
Source   &  \multicolumn{2}{c}{15~GHz} & Comments \\
         & Our result & MOJAVE result  &  \\
    & $m_c$ (\%)&  $m_c$ (\%) & \\
0109+224 & $-0.24\pm0.09$ & $<0.25$ & Our measurement consistent with upper limit\\
0133+476 & $-0.32\pm0.09$ & $-0.18\pm 0.09$ &\\
0300+470 & $-0.30\pm0.10$ & $<0.21$ &\\
0823+033 & $+0.20\pm0.09$ & $<0.26$ & Our measurement consistent with upper limit\\
0851+202 & $-0.15\pm0.08$ & $-0.20\pm 0.08$ & \\
         & $-0.19\pm0.08$ &  &\\
1055+018 & $+0.52\pm 0.10$ & $+0.32\pm0.09$  &\\
1156+295 & $-0.28\pm0.11$ & $-0.27\pm 0.09$  &\\
1253$-$055 & $+0.19\pm0.11$ & $+0.30\pm 0.08$ & \\
           & $+0.83\pm 0.10$ &  &\\
           & $+0.26\pm0.09$ &  &\\
1334$-$127 & $+0.28\pm0.09$ & $+0.29\pm 0.10$  &\\
1510$-$089 & $<0.22$ & $+0.20\pm 0.09$ & We find $+0.49\pm0.19$ at 22~GHz\\
1633+382 &   $-0.34\pm 0.06$ & $-0.39\pm 0.09$ & \\
1749+096 &   $-0.19\pm0.08$ & $<0.14$  &\\
         &   $-0.21\pm0.08$ &   &\\
2145+067 &   $-0.45\pm0.09$ & $<0.26$  &\\
2223$-$052 & $-0.20\pm0.07$ & $<0.22$ & Our measurement consistent with upper limit \\
2230+114 &   $-0.61\pm 0.08$ & $<0.19$ &\\
2251+158 &   $+0.17\pm0.10$ & $+0.23\pm0.10$ & \\
\hline
\end{tabular}
\end{table*}

We have detected CP in the jet, significantly displaced from the 
$I$ peak, in several sources, namely 3C279 (15, 22~GHz), 1334$-$127 (43~GHz) 
and possibly 1055+018 (43 GHz). We estimated the corresponding degrees of 
circular polarization using the total $V$ and $I$ fluxes in a 
$3\times 3$-pixel region around the $V$ peak. These values are listed in 
Table~\ref{tab:cp_noncore}, and are discussed below. 

\begin{table*}
\caption{CP Not Coincident with the VLBI Core} \centering
\label{tab:cp_noncore}
\begin{tabular}{lcclcc}
\hline
Source   &  Epoch & Frequency & Location & $m_c$ (\%) & $\sigma$\\
1055+018 & 15 March 2005 & 43~GHz & Northern/Southern side of jet & $-1.52/+0.87$ & 7.2/4.1\\
1253$-$055 &             & 15~GHz & Inner jet & +0.44 & 4.4  \\
           &             & 22~GHz & Inner jet & +0.79 & 3.1  \\
1334$-$127 & 04 March 2003 & 43~GHz & Western side of jet & $-7.16$ & 25.0 \\
\hline
\end{tabular}
\end{table*}

\subsection{Techniques Used to Test and Refine our Results}

The data in Tables~1 and 2 correspond to the CP maps we obtained using
the gain-transfer method of Homan and Wardle (1999). In several cases,
these images suggested that the $V$ peak was shifted from the total-intensity
($I$) peak. To test whether these shifts were real, we made a new $V$
image after applying one round of phase self-calibration assuming zero $V$, 
as described by Homan \& Lister (2006). If the CP signal in question is
not real, the apparent $V$ peak will tend to disappear from the new $V$ 
image.  If the CP in the image is real, 
but the shift from the $I$ phase center is due to phase errors, the  
peak of the new $V$ image should be nearly the same, but the $V$ peak
will now coincide with the $I$ peak. If the CP is real and its shifted
position from the $I$ phase center is also real, the effect of applying
the phase self-calibration assuming zero $V$ depends on the $I$ structure.
If the $I$ structure is extended, this additional phase self-calibration
may remove phase errors while leaving the fundamental information in the
$V$ phases intact, as was demonstrated by Homan \& Lister (2006). However,
if the $I$ structure is very compact, so that the $I$ phases are close to
zero, applying this phase-only self-calibration can corrupt the $V$
phases, giving rise to a false symmetrical $V$ structure about the $I$
phase center. Thus, the appearance of such a symmetrical structure upon
application of one phase self-calibration assuming zero $V$ to the
data obtained from the gain-transfer procedure can be used as a test of
the reality of the CP signal and its shift from the $I$ phase center.
We will refer to the application of one phase self-calibration assuming
zero $V$ as the PHC procedure.

%
%

\begin{figure*}
\centering
\includegraphics[width=0.45\textwidth]{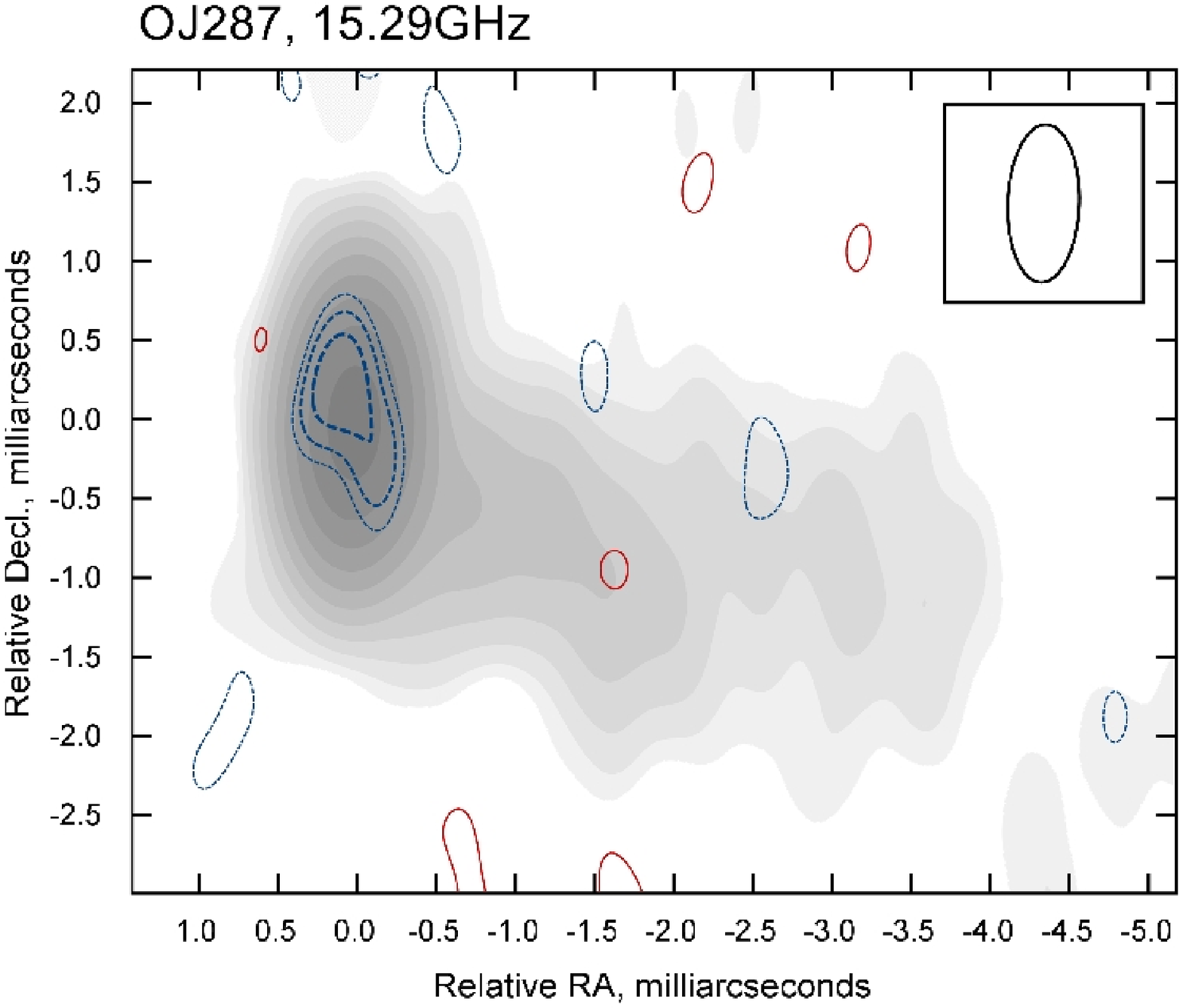}
\includegraphics[width=0.45\textwidth]{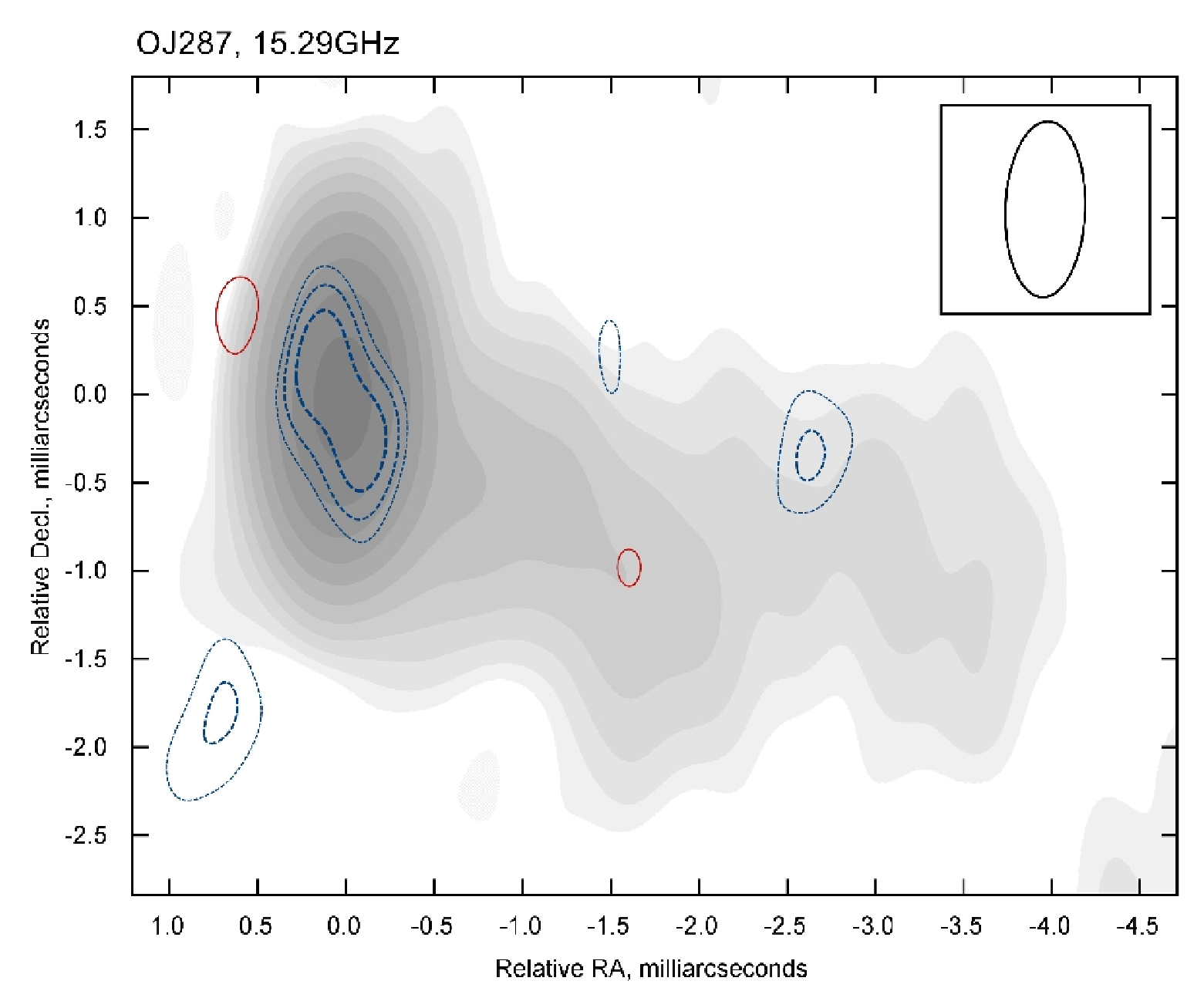}
\caption{Contours of $V$ superposed on a grey-scale representation of the
$I$ maps for OJ287 at epoch 26 September 2005. The two panels show the
gain-transfer $V$ map at 15~GHz (left) and the PHC $V$ map at 15~GHz
(right). The latter image clearly shows that the PHC procedure has 
artificially ``symmetrised'' the $V$ phases.}
\end{figure*}

\begin{figure*}
\centering
\includegraphics[width=0.45\textwidth]{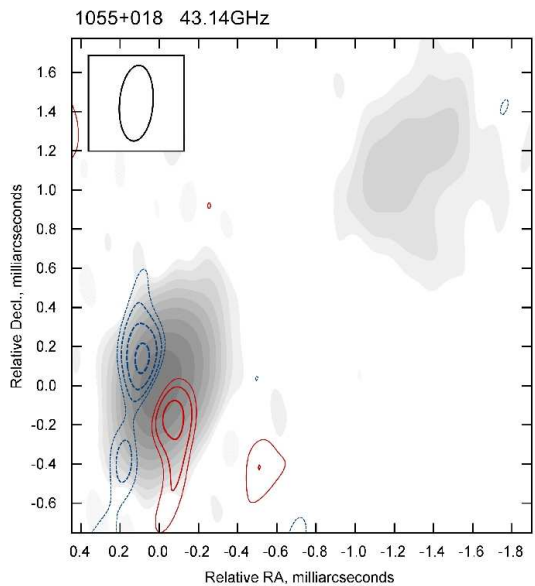}
\caption{Contours of $V$ for the gain-transfer map superposed on a 
grey-scale representation of the $I$ map for 1055+018 at 43~GHz 
at epoch 15 March 2005. 
}
\end{figure*}


\begin{table*}
\caption{Map Parameters for Figs.~1--5} 
\centering
\label{tab:maps}
\begin{tabular}{lccccccc}
\hline
Source   &  Epoch & Frequency & Procedure$^*$ & $V$ peak & Lowest $V$ contour & $I$ Peak & Lowest $I$ contour\\
0851+202 & 26 September 2005 & 15~GHz & GT & $-5.8$ & 2.2 & 3120 & 2.1 \\
         &                   &        & PHC& $-5.0$ & 1.9 & 3125 & 2.1\\
1055+018 & 15 March 2005     & 43~GHz & GT & $-20.0$& 6.3& 2261 & 5.0 \\
1253$-$055 & 04 March 2003   & 15~GHz & GT & +55.0& 4.6 & 6659 & 19.5 \\
           &                 & 22~GHz & GT & +44.9&17.0 & 7271 & 15.4 \\
           & 15 March 2005   & 43~GHz & GT & $-175.6$ & 41.1& 9571 & 25.1 \\
           &                 &        & PHC& $-93.2$ & 29.5 & 9478 & 24.6 \\
1334$-$127 & 05 March 2003   
                            & 43~GHz & GT & +13.3 & 5.9 & 1855 & 4.1\\
           &                 &        & PHC& +10.2 & 4.6 & 1857 & 4.0\\
2223$-$052 & 27 December 1999   & 15~GHz & GT & $-10.0$ & 2.6 & 5074 & 4.6 \\
\hline
\multicolumn{8}{l}{$^*$GT = gain-transfer; PHC = phase self-calibration assuming
zero $V$;}\\
\multicolumn{8}{l}{ RLC = separe $RR$ and $LL$ self-calibration.}
\end{tabular}
\end{table*}

Figs.~1--5 show $V$ images of the sources discussed below.  The gain-transfer 
$V$ images for all other objects in Table~\ref{tab:cp_core} show a single 
compact feature close to the $I$ phase center.
Table~\ref{tab:maps} lists the $V$ and $I$ peaks and lowest 
shown levels for the maps. The lowest $I$ contours correspond to twice the map 
rms and the contours increase in steps of a factor of two; the lowest $V$
contours correspond to 2.5 times the map rms, and the contours increase in
steps of a factor of $\sqrt{2}$. In each
figure, the convolving beam is shown in a corner of the maps.

{\bf OJ287 (0851+202).} Our 15-GHz gain-transfer $V$ map for epoch 
26 September 2005
is shown in Fig.~1 (left). The $V$ distribution is resolved in the 
direction of the jet, with the peak slightly displaced ``upstream'' from 
the $I$ peak.  When we apply the PHC procedure (Fig.~1, right), the 
amplitude of the $V$ peak is reduced, and a structure symmetrical about the 
$I$ phase center appears, as is expected if the true $V$ structure consists
of a single $V$ component that is shifted from the $I$ phase center. 
Formally speaking, we cannot be sure which peak in the GT image (upstream 
or downstream) corresponds to the real $V$ component; however, since the 
upstream peak is stronger, this more likely represents the 
real position of this component.  Accordingly, we conclude that
there is likely a small, real upstream displacement of the $V$ peak in
this source.

{\bf 1055+018.} Our 15 and 22-GHz gain-transfer $V$ maps for this AGN
both show point-like $V$ distributions that are well centered on the 
corresponding $I$ peaks. In contrast, our gain-transfer image at 43~GHz
shows regions of positive and negative CP placed on either side of the
innermost VLBI jet (Fig.~2). Spurious antisymmetrical structure of this
sort can appear {\em along} the VLBI jet in sources with extended structure,
due to slight misalignments of the $RR$ and $LL$ maps, as is discussed by
Homan \& Lister (2006). However, the $V$ structure observed in 1055+018 is
{\em across} the jet, and the $I$ structure is fairly compact at this
frequency. Unfortunately, we cannot test the reality of this structure
using the PHC procedure, because this procedure will lead to a cancelling
out of the two signals, even if they are real. Thus, we can make no
judgement about whether this structure is real or not, although we note
that it is somewhat difficult to understand what calibration errors would 
lead to a transverse misalignment between the $RR$ and $LL$ images. 
If this structure is real, the positive and negative peaks correspond to 
degrees of circular polarization $m_c = +1.5$ and $-0.9$,
which are within the range of the core CP values at 43~GHz in 
Table~\ref{tab:cp_core}.

\begin{figure*}
\centering
\includegraphics[width=0.37\textwidth]{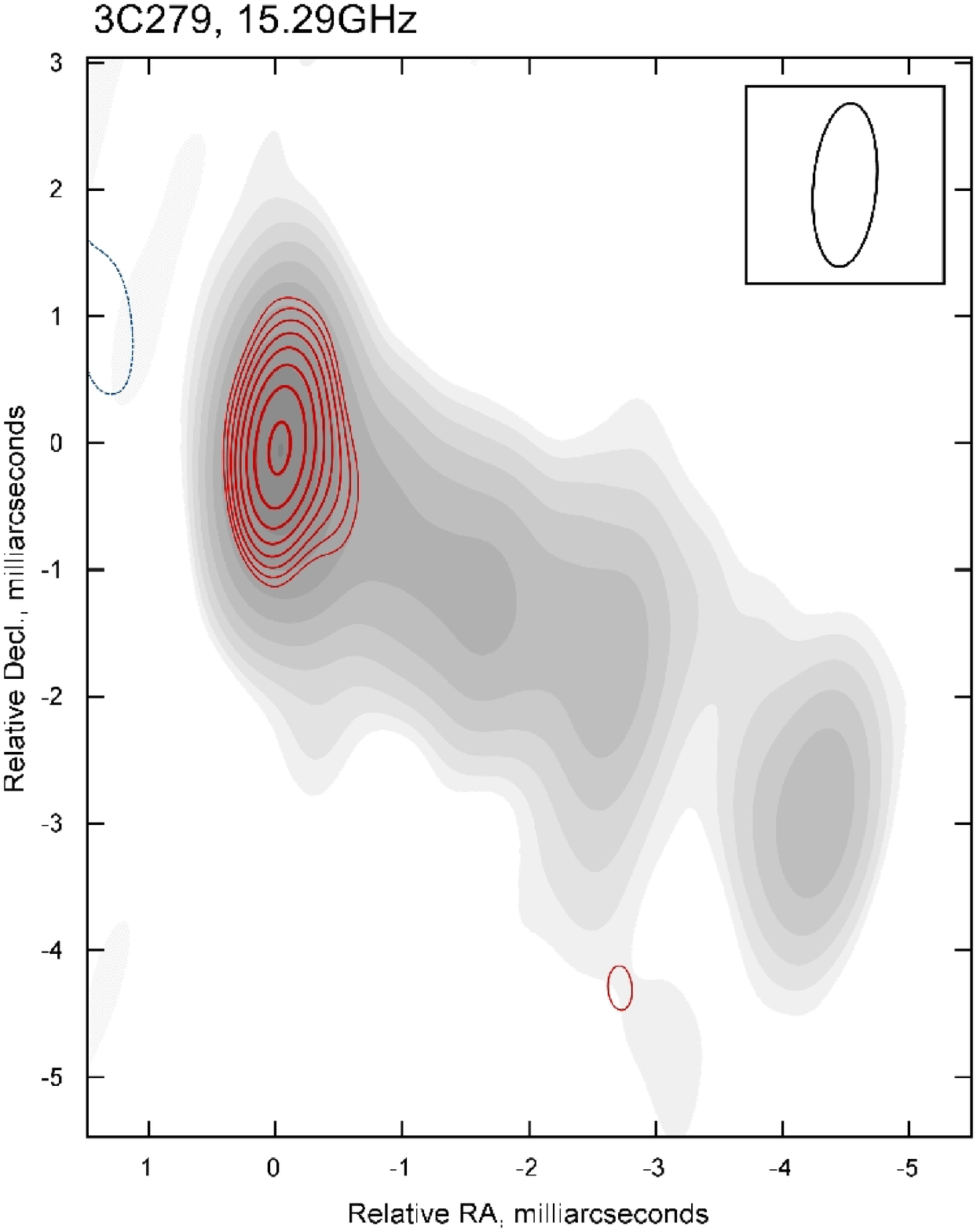}
\includegraphics[width=0.37\textwidth]{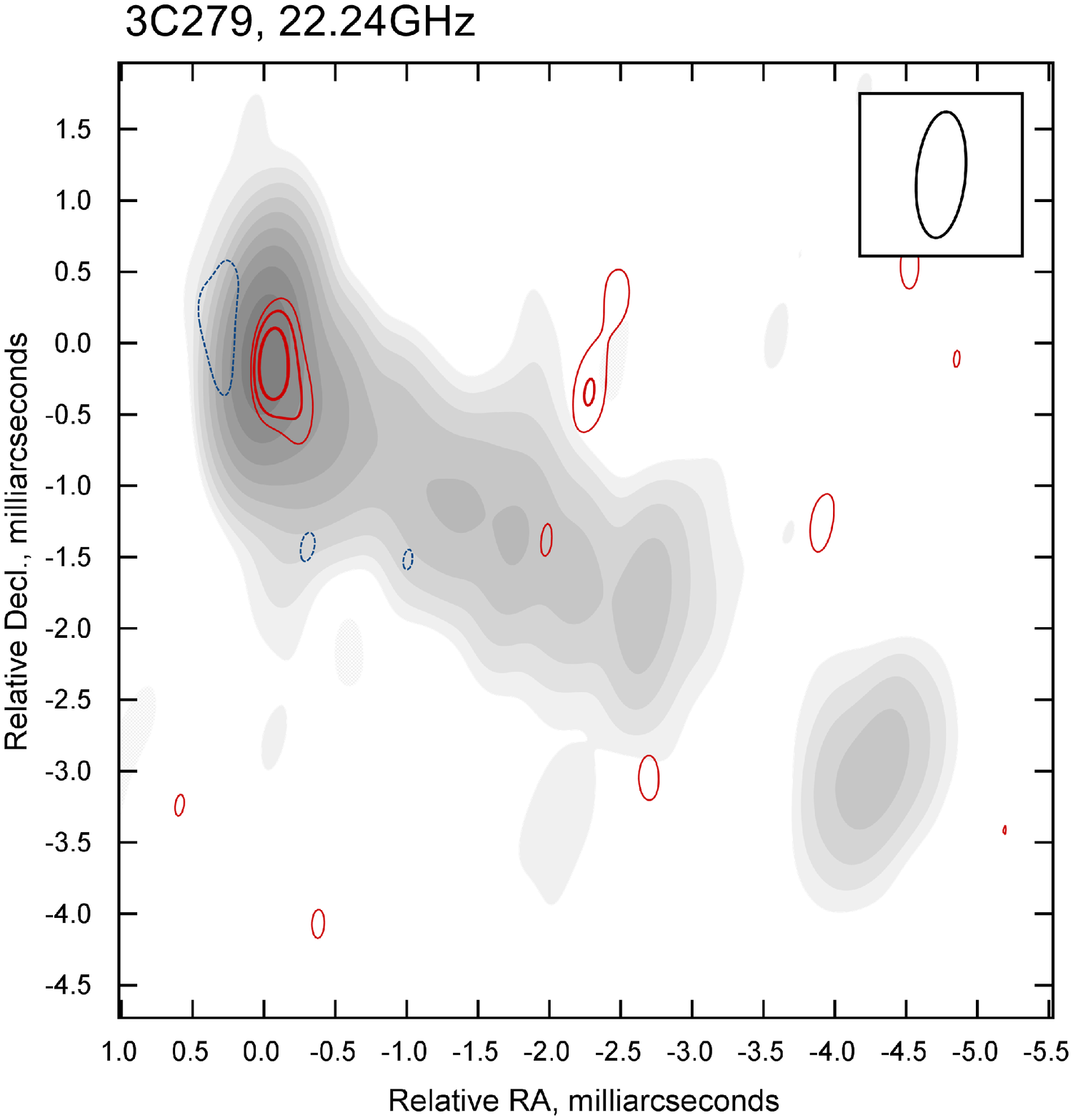}
\centering
\includegraphics[width=0.37\textwidth]{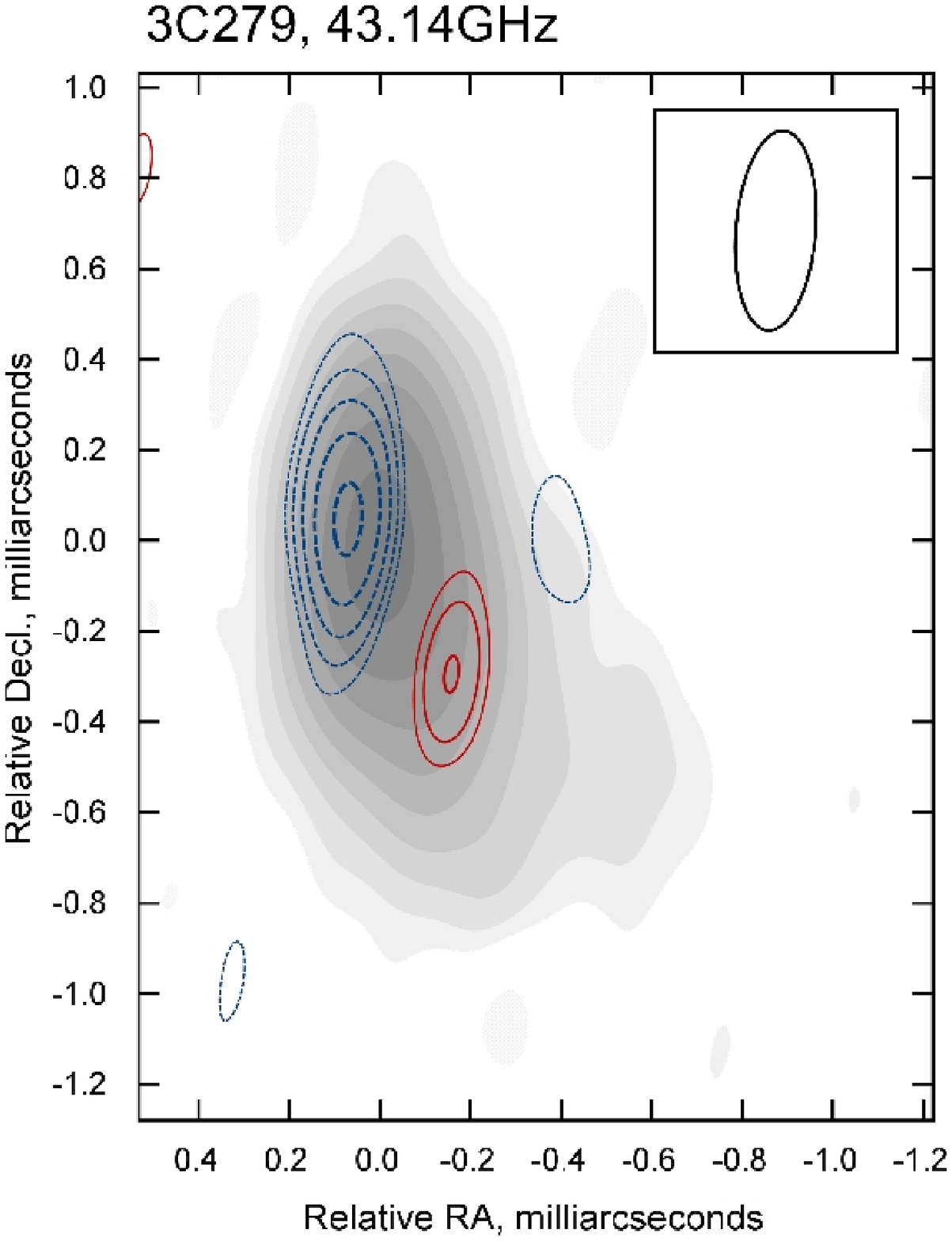}
\includegraphics[width=0.37\textwidth]{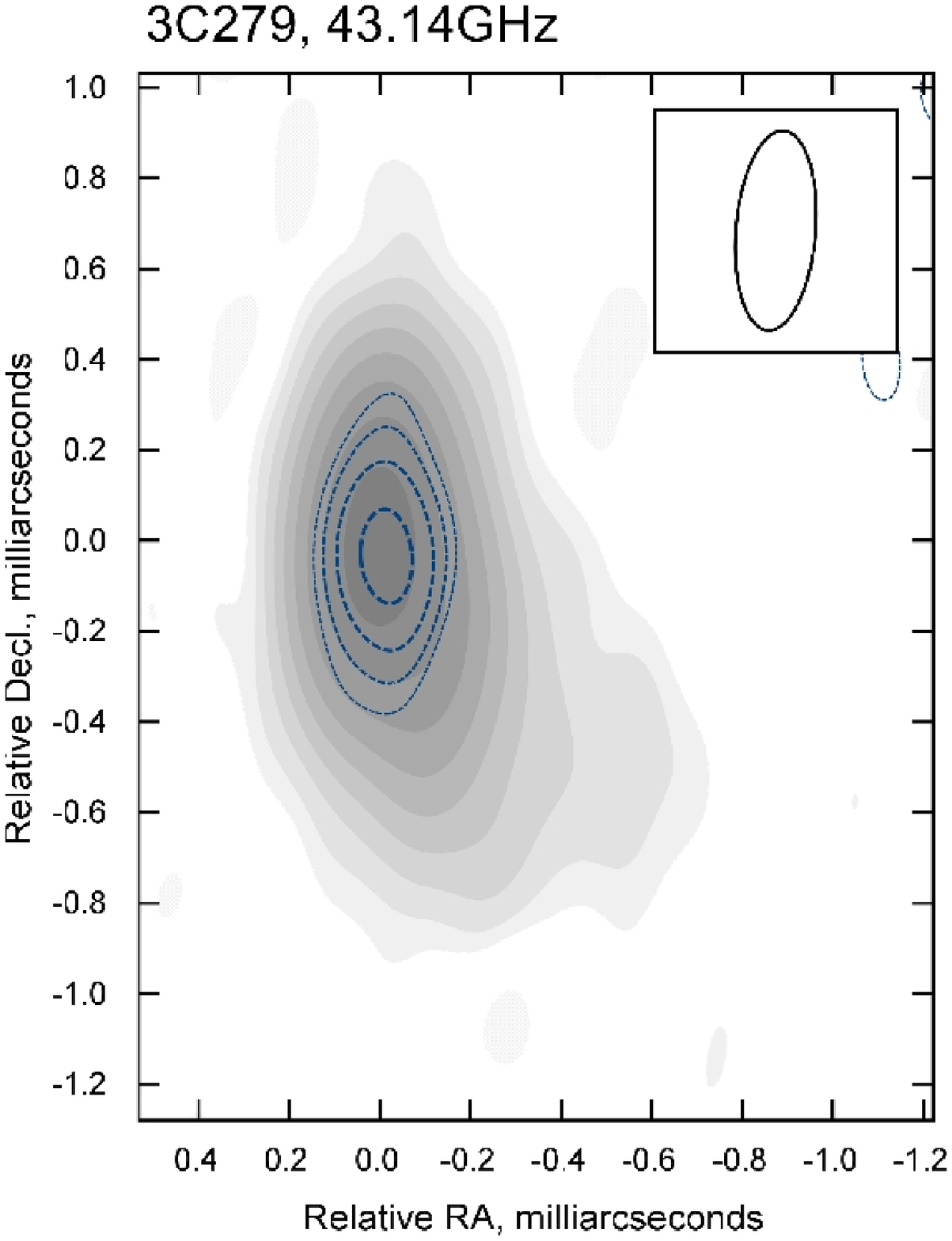}
\caption{Contours of $V$ superposed on a grey-scale representation of the
$I$ maps for 3C279.  The panels show the gain-transfer $V$ 
maps for epoch 04 March 2003 at 15~GHz (top left) and 22~GHz (top right), 
the gain-transfer $V$
map for 15 March 2005 at 43~GHz (bottom left) and the PHC $V$ map for
epoch 15 March 2005 at 43~GHz (bottom right).}
\end{figure*}

{\bf 3C279 (1253$-$055).} Our 15 and 22-GHz gain-transfer images for 3C279 
for epoch 04 March 2003 are dominated by a compact $V$ component that is 
essentially coincident with the $I$ peak, but in both, CP is also detected 
in the inner jet (Figs.~3, top).  We estimate the degree
of jet CP to be $m_c = +0.44$ and $+0.79$ at 15 and 22~GHz, within the
range of core values for $m_c$ at these frequencies.

3C279 is probably the single AGN for which the most CP measurements have
been made, most of them at 15~GHz. The sign of the 15-GHz CP has been 
consistently positive (Homan \& Wardle 1999, Homan \& Lister 2006, 
Table~1). Both of our 22-GHz measurements also yield positive CP. In
contrast, we detected weak positive CP at 43-GHz at epoch 14 March 2003, but
negative CP at 43-GHz at epoch 15 March 2005. The gain-transfer $V$ image for
this last dataset shows a negative component slightly upstream of the
$I$ peak and a weaker positive component out in the jet (Fig.~3, bottom
left).
This type of antisymmetric structure along the VLBI jet is consistent with
$RR$/$LL$ misalignments, and applying the PHC procedure 
leads to partial cancellation, leaving only a somewhat weaker negative 
$V$ peak that is well centered on the $I$ peak (Fig.~3, bottom right). 
We note, however, that this does not rule out the possibility  that the
structure displayed in the bottom left panel in Fig.~3 is real: such a
structure could come about if the ``core'' CP changed sign from positive
to negative, while the ``jet'' CP observed in March 2003 remained 
relatively stable.

\begin{figure*}[t]
\centering
\centering
\includegraphics[width=0.45\textwidth]{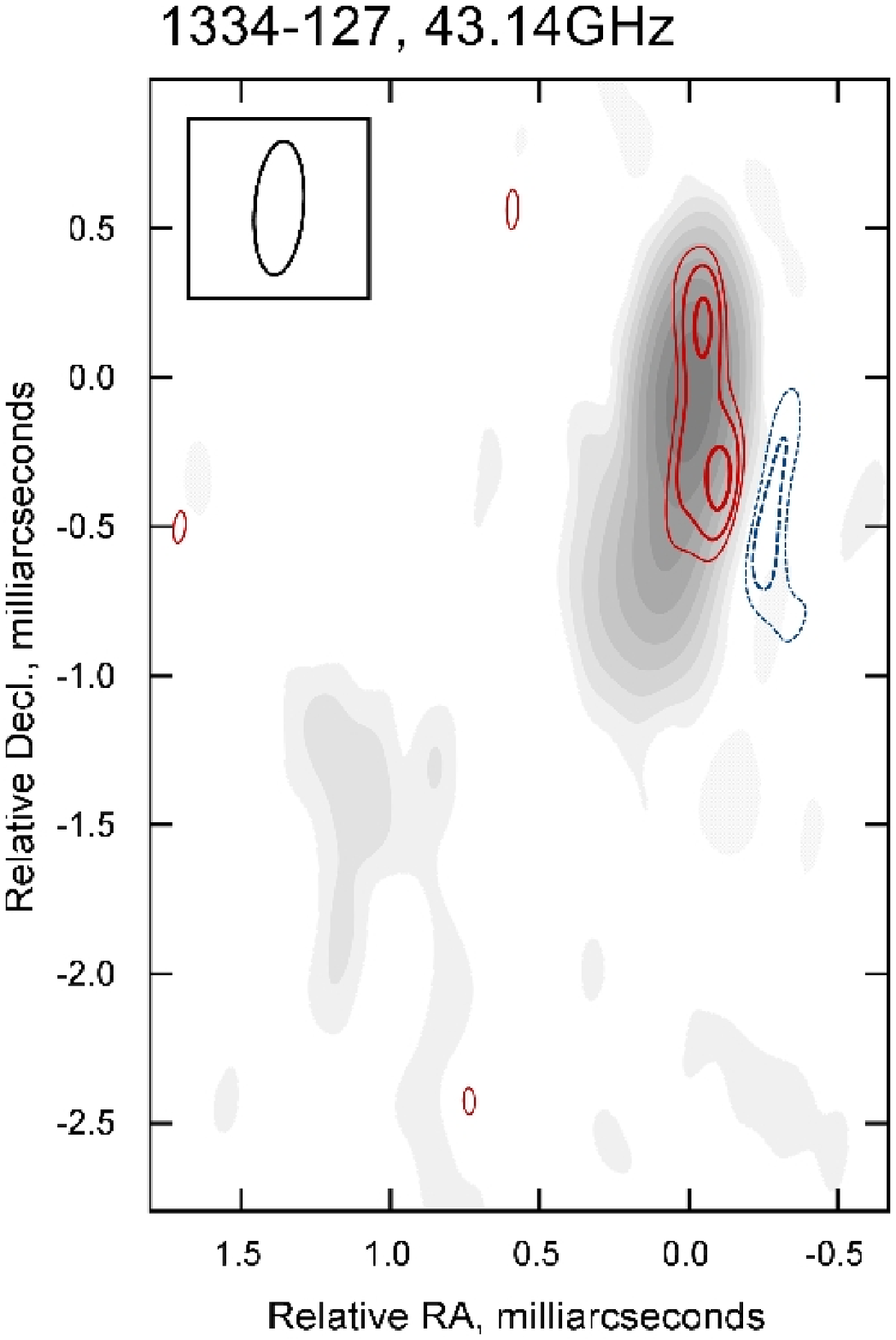}
\includegraphics[width=0.45\textwidth]{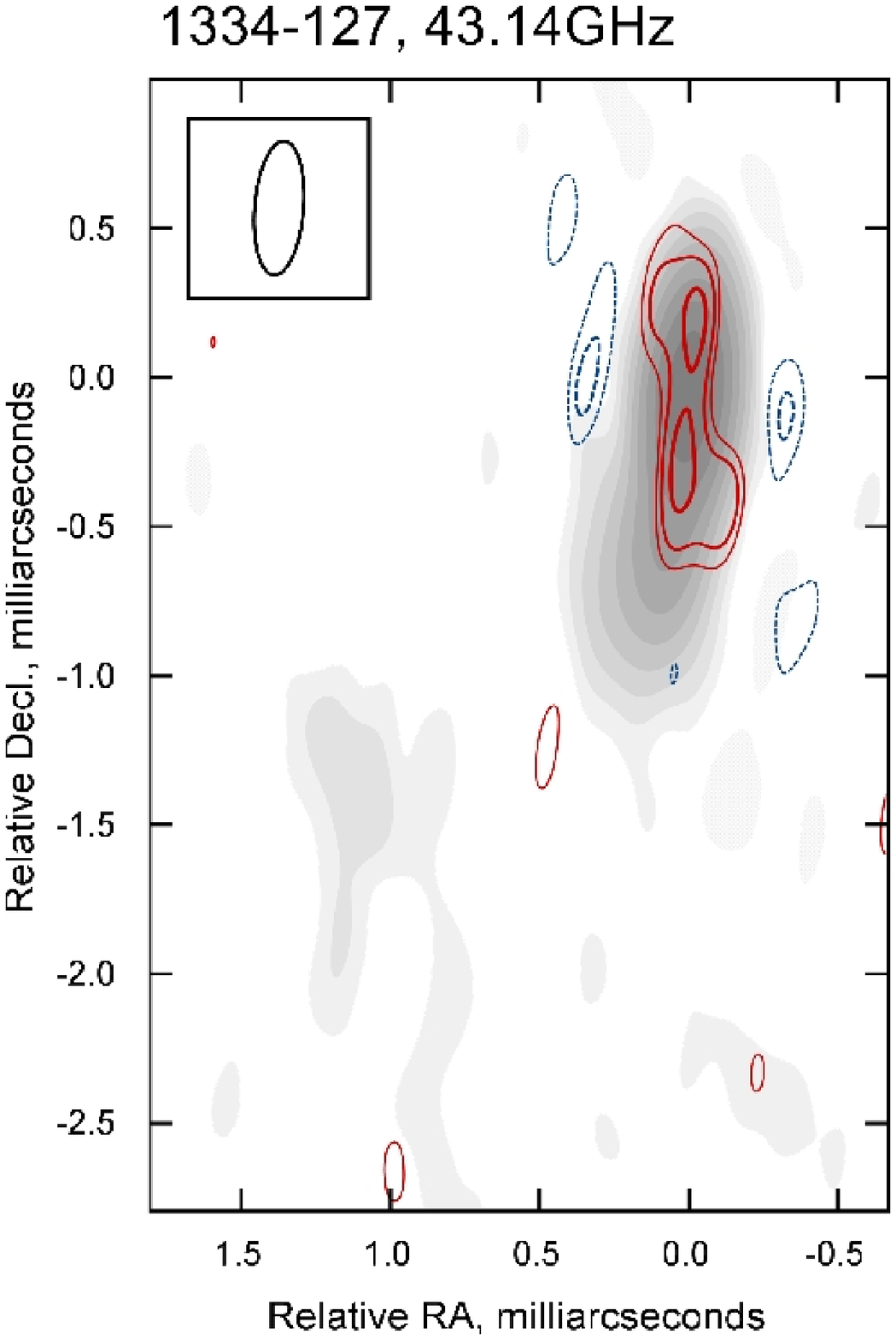}
\caption{Contours of $V$ superposed on a grey-scale representation of the
43~GHz $I$ map for 1334$-$127 for epoch 05 March 2003.  The panels show the 
43-GHz GT $V$ map (left) and the PHC $V$ map (right).} 
\end{figure*}

{\bf 1334$-$127.} The 15-GHz parsec-scale CP in this object seems very
stable, with the two measurements in Table~2 and the earlier measurement
of Vitrishchak \& Gabuzda (2007) all agreeing to within much less than
the quoted errors: $m_c = 0.28-0.29\%$. Our 15-GHz and 22-GHz GT 
$V$ images show compact emission well centered on the $I$ phase center. 
Our 43-GHz GT image shows a double 
structure with components with the same sign straddling the $I$ peak,
with the Southern component being somewhat stronger 
(Fig.~4, left). This structure is reminiscent of what we would 
expect if there were real CP offset from the $I$ peak, but the phases were 
not fully calibrated, and were overly biased toward values near zero. 
Application of the PHC procedure produces a very symmetrical 
structure about the $I$ phase center, consistent with the data containing 
a real CP signal that is shifted from the $I$ phase center 
(Fig.~4, right). 
Therefore, the 43-GHz CP detection is firm, and includes at least some
CP offset from the phase centre, probably to the South of the core (where
the peak of the GT map is located);  
however, we cannot be sure of the distribution of the CP, since residual phase 
errors may still be present in our GT $V$ image.  The inferred degree of 
CP for the peak of the CP distribution, which is located toward the
Southwestern edge of the jet, is $m_c = -7.16\%$. Although this is
high, it is only about a factor of two higher than the strong jet CP 
detected in 3C84 (Homan \& Wardle 2004). 

\begin{figure*}
\centering
\includegraphics[width=0.45\textwidth]{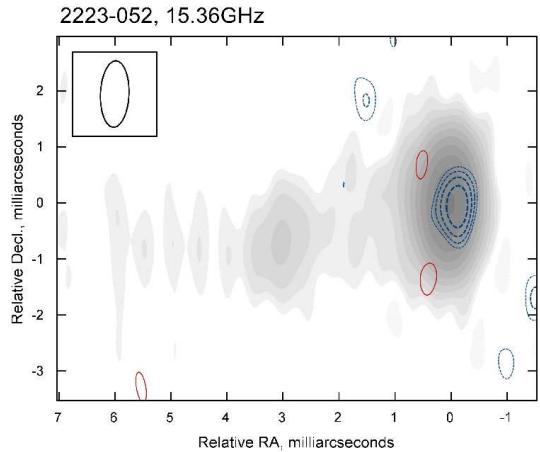}
\caption{Contours of the gain-transfer $V$ map superposed on a grey-scale 
representation of the 15-GHz $I$ map for 2223$-$052 for epoch 27 December 
1999.  
}
\end{figure*}

{\bf 2223$-$052.} Our 15-GHz gain-transfer $V$ image is resolved along
the jet direction, with its peak slightly upstream of the $I$ peak
(Fig.~5). The shape of the $V$ distribution mimics that
of the corresonding $I$ distribution, suggesting that both include
substantial contributions from inner jet emission that is close to the
observed core. 

\section{Discussion}

\subsection{Location of Detected CP in the VLBI Core Region and Inner Jet}

In most cases, the CP we have detected at all frequencies has been
coincident with the $I$ peak, which should be close to the position of
the VLBI core at these frequencies. This is also true of virtually all
previous CP measurements (Homan \& Wardle 1999, Homan \& Lister 2006,
Vitrishchak \& Gabuzda 2007). This has been taken as evidence that the
mechanism generating the CP operates efficiently near the base of the
jet, at the $\tau = 1$ surface, where $\tau$ is the optical depth (Homan \&
Wardle 1999).  However, it is important to note that the ``core'' as the 
optically thick base of the
jet (Blandford \& K\"onigl 1979) is a theoretical concept, and will correspond
to the observed ``core'' only for observations with sufficient resolution; 
the observed VLBI ``core'' will actually correspond to a combination of the 
genuine optically thick core and optically thin emission from the inner jet.
Thus, the detection of CP coincident with the observed VLBI ``core'' does
not necessarily mean that the $V$ signal arises predominantly near the
$\tau=1$ surface. 
This view is supported by the fact that jet CP has now been directly 
detected in nine AGN (Homan \& Lister 2006, Vitrishchak \& Gabuzda 2007,
the current paper).

At the same time, our 15 and 22-GHz CP images for OJ287 reveal $V$ peaks 
that are shifted slightly upstream of the $I$ peak, with the reality of these 
shifts
demonstrated by our tests using the PHC procedure. The position of the $I$ 
core is frequency-dependent, and will be shifted from the true jet origin in
the direction of the VLBI jet (e.g. Lobanov 1998), typically 
$\simeq 0.1-0.2$~mas at these frequencies. This suggests that the 
position of the $V$ peak is slightly closer to the true jet origin.
In general, the location of the $V$ peak will depend on magnetic-field 
strength and electron density; one possibility is that the appearance of
the $V$ peak slightly upstream of the $I$ peak is associated with the
imminent birth of a new VLBI component. There is some support for this
from a comparison of our 15-GHz linear-polarization map with the MOJAVE
linear polarization maps for epochs shortly before and after our own. In
particular, the $I$ peaks for the two MOJAVE maps preceeding our own 15-GHz 
observations were 2200~mJy (02 December 2004) and 2040~mJy (03 June 2005),
with the core polarization angle close to orthogonal to the jet direction,
which lies in position angle $PA\simeq -120$; at our epoch roughly four
months later (26 September 2005), the core polarization angle retains this 
same orientation, but the $I$ peak has increased to 3120~mJy; by the 
following MOJAVE epoch (28 April 2006), the $I$ peak had increased to 
4570~mJy, and the core polarizaton angle had rotated to a nearly vertical 
position. By 06 January 2007, the MOJAVE $I$ peak had gone back down to 
2340~mJy. Thus, it is clear that the core was in the process of undergoing 
an outburst during our VLBA epoch; we can speculate that the location of the 
$V$ peak slightly upstream of the $I$ peak may have been related to this 
outburst, but time series of CP measurements are needed to determine the
likelihood of this scenario.

The idea that the CP detected in the observed VLBI ``core'' may often
be associated with (partially) optically thin regions in the innermost jet 
is consistent with CP generated by either the synchrotron mechanism or the
Faraday conversion of linear to circular polarization. 
In either case, the CP can be generated to some extent on all scales 
in the jets, but it is natural that the innermost jet (within the observed
VLBI ``core'') dominates the CP signal, due to the strong {\bf B} fields
and plasma densities in this region.  We have argued recently that the 
CP is likely generated by Faraday conversion in helical jet magnetic 
fields, with linearly polarized synchrotron emission radiated at the far 
side of the jet relative to the observer being partially converted to 
circular polarization in the near side of the jet (Gabuzda et al. 2008).  

\begin{table*}
\caption{CP spectra ($|m_c| \propto \nu^{\alpha_c})^{\dagger}$} \centering
\label{tab:indices}
\begin{tabular}{lccccccccc}
\hline
Source   &  15~GHz $m_c$& 22 GHz $m_c$ & 43 GHz $m_c$ & $\alpha_{c}$ & $\alpha_c$ & $\alpha_c$ & $\alpha_I$ & $\alpha_I$ & $\alpha_I$\\
    & (\%)& (\%)&  (\%)& 15--22 & 22--43 & 15--43 & 15--22 & 22--43 & 15--43\\
0133+476 &  $-0.32\pm0.09$ & -- & $-0.33\pm0.19$ & --  & --  & $+0.030\pm0.64$ & $-0.03$& +0.11& +0.00\\
0851+202 & $-0.19\pm0.08$ & $-0.20\pm0.13$ & $+0.55\pm0.26$ & $+0.14\pm 0.77$ &$+1.53\pm0.80$ & $+1.02\pm 0.63$ & +0.12 & $-0.09$ & $-0.01$\\
1055+018 & $+0.52\pm 0.10$ & $+0.27\pm0.17$& * & $-1.75\pm0.66$ & -- & --  & +0.15 & $-0.39$ & $-0.19$\\
1253$-$055 & $+0.83\pm 0.10$ & $+0.62\pm0.25$ & $+1.21\pm0.37$ & $-0.78\pm0.42$ & $+1.01\pm 0.51$ & $+0.36\pm 0.33$ & +0.37 & $-0.75$ & $-0.35$\\
& $+0.26\pm0.09$ & $+0.20\pm0.15$ & $-1.03\pm 0.16$ & $-0.70\pm 0.83$ & $+2.47\pm 0.77$ & $+1.32\pm 0.38$ & +0.16 & $-0.02$ & +0.04\\
1334$-$127 & $+0.28\pm0.09$ & $+0.40\pm0.24$& * & $+0.95\pm 0.68$ & -- &  --& $-0.14$ & $-0.76$ & $-0.54$\\
1510$-$089 & -- & $+0.44\pm0.19$& $-2.43\pm0.40$ & -- & $+2.58\pm 0.46$ & -- & $-0.18$ & $-0.40$ & $-0.34$\\
1633+382 & $-0.34\pm 0.06$ & $-0.83\pm 0.17$& -- & $+2.38\pm 0.27$ & -- & -- &+0.09 & $-0.32$ &$-0.17$\\
2145+067 & $-0.45\pm0.09$  & $-0.34\pm0.13$ & -- & $-0.75\pm 0.43$ & -- & -- & $-0.20$& $-0.36$ &$-0.30$\\
2230+114 & $-0.61\pm 0.08$ & $-1.26\pm0.21$ & -- & $+1.94\pm 0.21$ & -- & -- &+0.48& +0.46 &+0.48\\
\hline
\multicolumn{7}{l}{$^{\dagger}$All $m_c$, $\alpha_c$ and $\alpha_I$ values determined after 
convolution with 15~GHz beam.}\\
\multicolumn{7}{l}{$^{*}$CP possibly detected off core.}
\end{tabular}
\end{table*}

\subsection{Spectra of the Degree of Polarization}

We can define spectral indices for the degree of CP, 
$\alpha_c$ ($|m_c|\propto\nu^{\alpha_c}$), and total intensity, 
$\alpha_I$ ($I\propto\nu^{\alpha_I}$). Table~\ref{tab:indices}
presents $m_c$, $\alpha_c$ and $\alpha_I$ for the cores in which CP
was detected. We are using $\alpha_c$ as an indicator of the frequency
dependence of the degree of CP, independent of the CP sign, and so consider
only the magnitude of $m_c$ when determining $\alpha_c$. In this sense,
$\alpha_c$ characterises the frequency dependence of the efficiency of the
CP generation. 

There is no clear universal trend for the frequency dependence of the 
degree of circular
polarization, although a few patterns can be noted. When CP is detected 
at 43~GHz, the degree of CP at this frequency is higher than at the lower 
frequencies. In other words, all the values for $\alpha_c$ between 22 and 
43~GHz and between 15 and 43~GHz are positive. We are confident that this
result is not due to observational effects. The average value for $|m_c|$
for the 15-GHz measurements in Table~5 is $0.42 \pm 0.03$, while the average 
$|m_c|$ at 22~GHz is $0.49\pm 0.06$, and the average $|m_c|$ at 43~GHz is
$1.00\pm 0.01$. This shows clearly that the average degrees of CP at
15 and 22 GHz essentially coincide, while the average degree of CP at
43~GHz is appreciably larger. This demonstrates that, treated as a group,
as well as on a source-by-source basis, the 43-GHz $|m_c|$ values are
systematically higher than the $|m_c|$ values at the lower two frequencies. 
In contrast, roughly half the $\alpha_c$ values between 15 and 22~GHz are 
positive and half are negative; i.e., the $|m_c|$ values are as often
higher as lower at 22~GHz compared to 15~GHz. 
 
The ranges for several of the 15--22~GHz $\alpha_c$ values encompass 
the nominal spectral index expected for intrinsic CP 
from a homogeneous source in the optically thin regime, 
$m_c\propto\nu^{-0.5}$. However, given the large range of observed 
15--22-GHz $\alpha_c$ values (from +2.38 to $-1.75$), this may well 
be a coincidence, particularly since the associated errors are fairly 
large, and all the cores have intensity spectral indices indicating that 
they are partially optically thick.  The nominal spectrum expected for CP 
generated by Faraday conversion in a homogeneous source is substantially 
steeper than any of the observed negative CP spectral indices, 
$m_c\propto\nu^{-3}$. 

One way in which it might be possible to obtain a wide range of CP spectral 
indices for the core region is if there are several regions of CP
contributing to the observed 
``core'' CP. If we are observing the sum of two or more CP components, 
possibly having different signs and somewhat different spectra, this
could lead to a fairly wide range of both positive and negative values for
$\alpha_c$, as is observed.  There is some basis for this type of picture. 
For example, Homan \& Wardle (2004)
found very different $\alpha_c$ values for three CP components 
in the inner jet of the nearby ($z = 0.017$) radio galaxy 3C84: 
$\alpha_c = -0.9$ for a component with positive CP in
a predominantly optically thin region and $\alpha_c = +1.4$ and +1.7 for 
two components with negative CP in partially optically thick regions. 
These three regions would all be within the observed ``core'' 
if the galaxy were at a redshift more typical of the AGN considered
here, and it is possible that we are seeing precisely the effect of such
a mixing of contributions from different components in our ``core'' CP
spectra. Both intrinsic CP and CP generated by Faraday conversion 
in a helical {\bf B}-field geometry can also have comparably
strong regions of CP of different signs in different regions of the jet
(e.g. on different sides of the jet; see Fig.~6 of Gabuzda et al. 2008), 
which could be blended in the observed ``core.'' 

It is also quite possible -- even likely -- that the frequency dependence 
for the observed CP is associated to a considerable extent with the 
intrinsic inhomogeneity of the jets. Wardle \& 
Homan (2003) discuss the approximate frequency dependences that might
be expected for a Blandford--K\"onigl (1979) jet under various conditions.
Such a jet is essentially self-similar, with the magnetic field and electron
density falling off with distance from the jet base as $B\propto r^{-1}$
and $n_e\propto r^{-2}$. This leads to the expectation that, like the
observed core total intensity, intrinsic CP from the inner 
jet should have a roughly flat spectrum; scaling arguments suggest that 
the expected spectrum for CP generated by Faraday conversion driven by 
Faraday rotation is also roughly flat. On the other hand, if the unidirectional
magnetic field falls off as $r^{-2}$, which conserves the poloidal magnetic
flux, scaling arguments suggest that both intrinsic CP and CP due to 
Faraday-rotation-driven conversion will have an {\em inverted} spectrum, 
$m_c\propto \nu^{+1}$.  Physically, this essentially comes about due to
the characteristic fall-offs in $B$ and $£n_e$ with distance from the 
jet base, combined with the shift of the location of the $\tau=1$ 
surface closer to the true jet base with increasing frequency, assuming that
the dominant contribution to the CP is made by regions roughly in the 
vicinity of the $\tau=1$ surface. 

The expected frequency dependence for the case of CP generated by
Faraday conversion in a helical {\bf B} field associated with a
Blandford--K\"onigl jet is not obvious. Independent of frequency, 
the degree of CP will depend on the pitch angle, with the conversion 
being most efficient for pitch angles near $\psi = 22.5^{\circ}$ and
$67.5^{\circ}$ (i.e. for angles between the ``background'' and 
``foreground'' components of the helical {\bf B} field being 
$2\psi = 45^{\circ}$ or $135^{\circ}$), if the internal
Faraday rotation arising in the jet is not too large (Gabuzda et al. 2008).
If internal Faraday rotation is substantial,
this could either increase or decrease the amount of Faraday conversion,
depending on the pitch angle of the jet and whether the Faraday rotation
drives the angle between the plane of polarisation for the synchrotron 
radiation emitted at the back of the jet and the {\bf B} field at the
front of the jet toward or away from $45^{\circ}$ (or $135^{\circ}$).
Since, as in the cases of intrinsic CP and CP due to 
Faraday-rotation-driven conversion, helical-field-driven conversion is governed
by the ordered {\bf B} field, which we suppose to fall off with distance 
from the jet base more rapidly than the disordered field component, 
it seems plausible that similar scaling arguments can be applied for the
CP generated in this case, giving rise to an inverted spectrum for CP 
generated roughly in the vicinity of the $\tau=1$ surface.

In a nutshell, what this means is that it is not easy to definitively
distinguish between the synchrotron mechanism and either 
Faraday-rotation-driven or 
helical-field-driven Faraday conversion based purely on the observed
$m_c$ spectra. This leaves the observed degrees of CP as a possible means
of constraining the mechanism at work. It has been argued in previous
studies (e.g. Homan \& Wardle 2004) that intrinsic CP 
has trouble explaining the highest observed degrees of circular 
polarization.  The degree of intrinsic CP is

\begin{eqnarray*}
m_c & = & \epsilon_{\alpha}^{\nu}\left(\frac{\nu_{B\perp}}{\nu}\right)^{0.5}
\frac{B_{u,los}}{B_{\perp}^{rms}}
\end{eqnarray*}

\noindent
where $\epsilon_{\alpha}^{\nu}$ is a constant that has values ranging
roughly from 0.45--1.40 for $-0.25 \leq \alpha\leq +0.25$ (as is observed 
for most of our cores); $B_{u,los}$ is the component of the 
uniform magnetic field that is responsible for generating 
the circular polarization; $B_{\perp}^{rms}$ is the mean field component 
in the plane of the sky,
which includes both the transverse part of the total uniform magnetic field
$B_u$ and any disordered field
that contributes to $I$ but not $V$; and $\nu_{B\perp} = 2.8B_{\perp}^{rms}$
is the gyrofrequency for $B_{\perp}^{rms}$ in MHz (Jones \& O'Dell 1977;
Wardle \& Homan 2003). Assuming as fairly
reasonable and typical values for AGN with relativistic jets
and $B_{u,los}/B_{\perp}^{rms}\approx 0.10$, we can
obtain $m_c$ values up to about 0.40\%, 0.30\% and 0.25\% at 15, 22 and
43~GHz, respectively, for $B_{\perp}^{rms} \approx 0.40$~G, consistent with 
estimates of core magnetic fields derived from the frequency-dependent 
core shift (Lobanov 1998, O'Sullivan \& Gabuzda, in prep.). Thus, while 
some of the observed
$m_c$ values could, in principle, potentially be generated by the synchrotron 
mechanism, an appreciable number of the observed degrees of CP at all 
three of our frequencies would require either magnetic fields that were 
implausibly high, $B_u\simeq 2-60$~G, or fields that were implausibly well 
ordered, $B_u/B_{perp}^{rms}\approx 0.5-1.5$. The very high degree of CP
observed for the VLBI core of 1510$-$089 at 43~GHz, $m_c \simeq -2.8\%$
(Table~1), is particularly problematic for the synchrotron mechanism.

Overall, it seems likely that the observed CP is predominantly
generated by Faraday conversion, 
and that the observed values of $\alpha_c$ are determined 
by effects associated with the intrinsic inhomogeneity of the jets, 
as well as the possible presence of several regions of CP of either 
one or both signs contributing to the observed ``core'' CP. Based on
the arguments presented by Gabuzda et al. (2008), we suggest that the
predominant mechanism generating the observed CP is helical-field-driven
Faraday conversion. 

\subsection{Sign Changes with Frequency}

It has been known for some time that the sign of the CP in a particular
AGN is usually stable over several years, or even decades (Homan \& Wardle
1999). In contrast, little has been known about the frequency dependence
of the CP sign.  We have detected CP at more than one frequency in 10 AGN.
Of the 9 AGN for which CP is detected at both 15 and 22~GHz, 8 show the
same sign at these two frequencies, the exception being 2251+158. 
Of the 6 AGN for which CP was
detected at both 22 and 43~GHz, 4 show sign changes between these two
frequencies (0851+202, 1253$-$055, 1510$-$089, 2251+158).
The appearance of more frequency CP sign changes in the transition between
22 and 43 GHz appears to suggest optical depth effects; i.e., that the
regions being sampled are optically thin between 22 and 43~GHz, but
optically thick between 15 and 22~GHz. 
However, there is no evidence
that this transition occurs in our frequency range from the $I$ spectral
indices (Table~5), which are similar between 15--22~GHz and 22--43~GHz.
The $\alpha_I$ values for these two frequency ranges for
0851+202 (OJ287) are +0.12 and $-0.09$; for 1253$-$055 (3C279) are +0.16 
and $-0.02$; for 1510$-$089 are $-0.18$ and $-0.40$; and for 2251+158 are
$+1.30$ and $+1.28$. 

Previous measurements of the VLBI-core CP of 3C279 at 15~GHz have all
showed it be positive (Homan \& Wardle 1999, Homan \& Lister 2006). Our
data likewise show the 15~GHz CP of 3C279 to be positive at epoch
07 August 2002, and the CP at all three frequencies at epoch 04 March 2003. 
In contrast, the 43-GHz CP had changed to negative two years later, at
epoch 15 March 2005.  This was accompanied by a substantial weakening of the 
degree of CP observed at 15 and 22~GHz, suggestive of the growth
of a component with negative CP that partially cancelled the previously
present positive CP component at the two lower frequencies, and already 
dominated at the highest frequency. This behaviour could be due to the
frequency dependences of the two components (positive and negative), or
alternatively to the spatial scales on which they are present. According
to the MOJAVE 15-GHz polarization images,
this possible development of a negative CP component was
accompanied by a substantial brightening of the 15-GHz VLBI core: the
15-GHz $I$ peak rose from 9540~mJy/beam on 11 June 2004 to 10,380~mJy/beam on
05 March 2005, then to 12,440~mJy/beam on 15 June 2005, subsequently falling 
to 10,990~mJy/beam by mid-September 2005.  The 43-GHz VLBA monitoring of
the Boston University group confirms that there was an outburst in the 
VLBA core of 3C279 peaking in early 2005, which appears to have been
associated with the birth of a new VLBI component in late 2004 (Chatterjee
et al., in preparation). This suggests that the inferred
appearance of a negative CP component may have been associated with the
emergence of this new VLBI component, but this is only speculative at
present, due to the limited observational information available.

\subsection{Transverse CP Structures}

We have found tentative evidence for transverse CP structures in 1055+018 and 
1334$-$127 at 43~GHz. 
Our tests of the reality of the off-core CP in 1334$-$127 using the PHC
procedure provide support for the presence of off-core CP, probably 
slightly South of the core and displaced toward the edge of the jet.
The PHC test is unfortunately not suitable for the 
transverse CP structure in 1055+018, since it consists of features
with quite similar intensities and opposite signs located on opposite
sides of the jet, nearly symmetrically about the core, and so could come 
about due to a slight misalignment between the $RR$ and $LL$ images.  We 
argue above that it is much more natural for such misalignments to come 
about in the direction of the jet rather than orthogonal to it, and
for this reason we have kept 1055+018 on our list of sources possibly
displaying  transverse CP structure. 


\begin{figure*}
\centering
\includegraphics[width=0.48\textwidth]{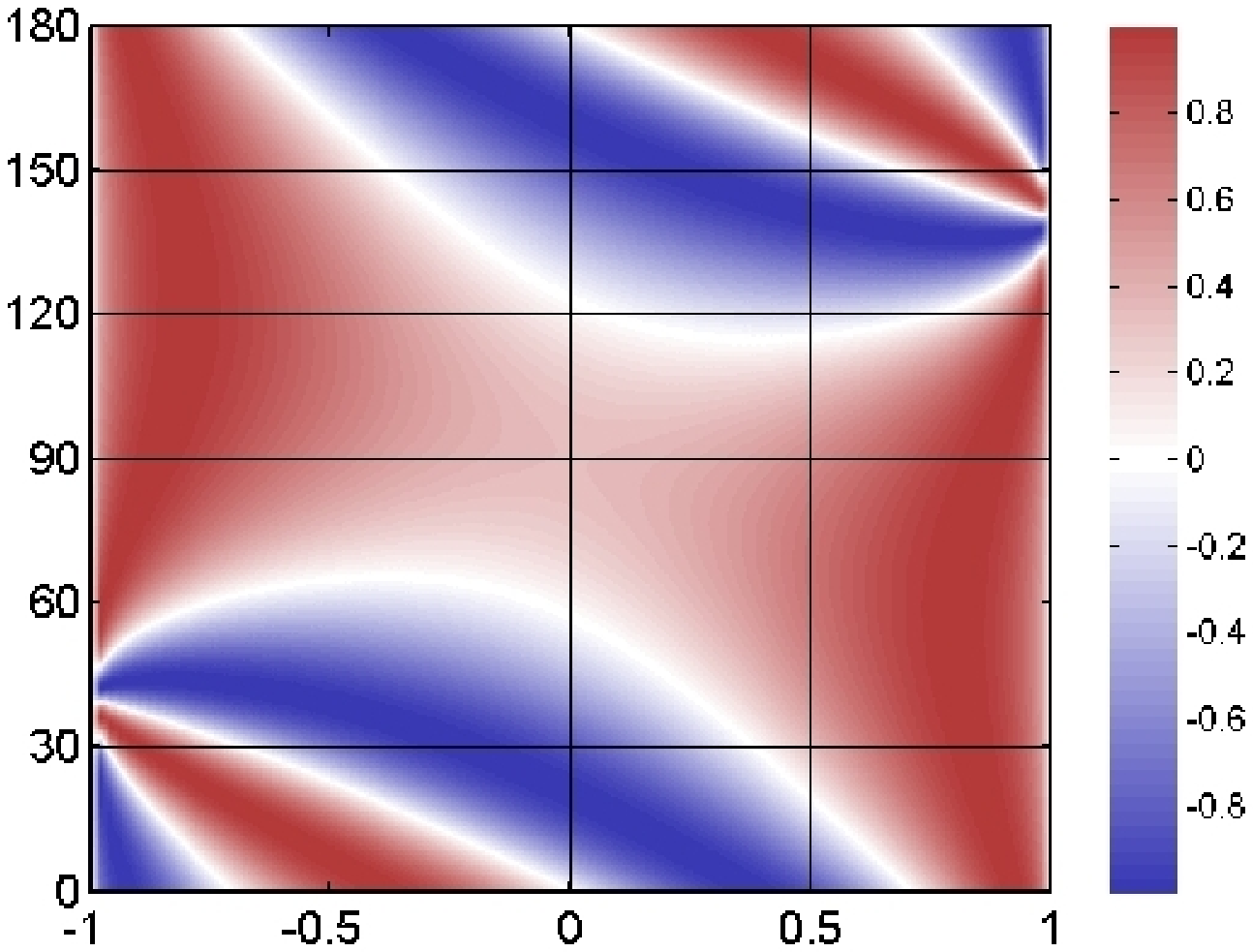}
\includegraphics[width=0.48\textwidth]{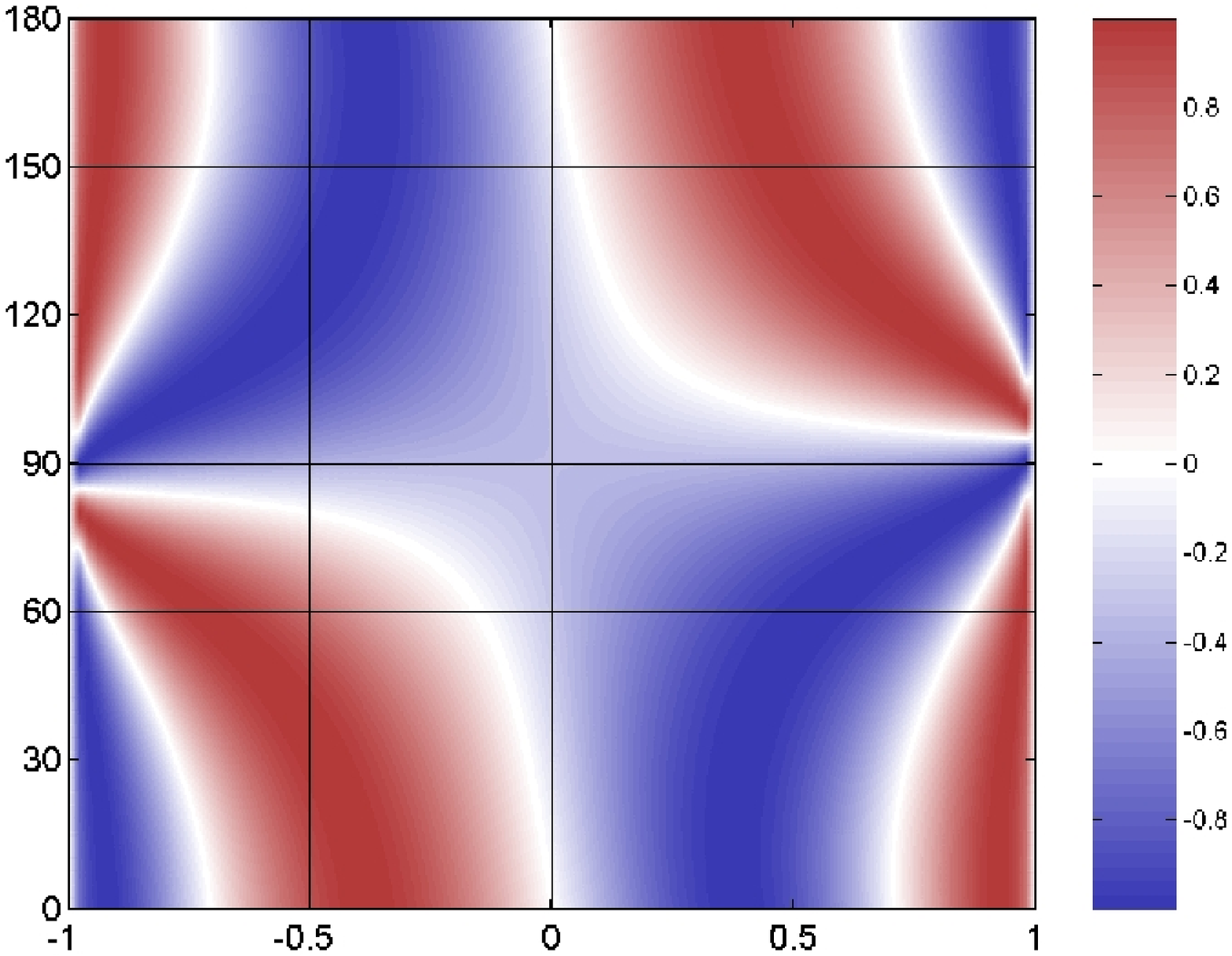}
\centering
\caption{Diagrams showing the sign of CP generated by Faraday
conversion in a helical jet {\bf B} field as a function of the relative
location across the jet (horizontal axis running from $-1$ at left to $+1$
at right, 0 corresponds to the ``spine'' of the jet) and the angle at which
the jet is viewed in the jet rest frame
(vertical axis running from $0^{\circ}$ at bottom to $180^{\circ}$ at top;
$0^{\circ}, 90^{\circ}$ and $180^{\circ}$ correspond to an observer viewing
the jet ``head-on'', ``side-on'' and ``tail-on'', respectively). The plots
shown are for right-handed 
helical fields with pitch angles of $40^{\circ}$ (top) and $85^{\circ}$ 
(bottom). Relatively low pitch angles give rise to a single dominant 
region of a single CP sign all across the jet, which is strongest near 
one edge of the jet for jet rest-frame viewing angles not too far from
$90^{\circ}$, while high pitch angles give rise to regions of positive 
and negative CP on opposite sides of the jet for jet rest-frame viewing 
angles not too far from $90^{\circ}$.}
\end{figure*}

One reason we feel behooved to include our 43-GHz CP map for 1055+018
in this paper, even though we cannot firmly demonstrate that the observed
structure is real, is that transverse CP structures such as those our 
analysis suggest for 1055+018 and 1334$-$127 would come about naturally if 
the CP is generated in helical jet {\bf B} fields, independent of whether
the CP is due to the synchrotron mechanism or helical-field-driven
Faraday conversion.
Gabuzda et al. (2008) considered a simple model in which the CP in AGNs is 
generated by the latter mechanism; the sign of the 
CP is determined essentially by the angle between the {\bf B} fields at 
the far side and near side of the jet relative to the observer, which 
changes across the jet.  Whether the CP is dominated by a single sign 
throughout the jet cross section or includes contributions of both positive 
and negative CP on opposite sides of the jet is determined by the pitch 
angle and viewing angle. This is illustrated in Fig.~6, which shows the
sign of CP generated by Faraday conversion in a helical jet {\bf B} field 
as a function of the relative location across the jet and the angle at which
the jet is viewed in the jet rest frame. A viewing angle of $\simeq 
1/\gamma$ in the observer's frame corresponds to a viewing angle of
$90^{\circ}$ in the jet rest frame, where $\gamma$ is the jet Lorentz
factor; the most likely viewing angle has been estimated to be slightly
smaller than this, $\simeq 1/2\gamma$, which corresponds to a viewing
angle in the jet rest frame of about $53^{\circ}$ for Lorentz factors
thought to be typical of the VLBI jets ($\gamma\sim 10$). Thus, the jets of 
core-dominated radio-loud AGN are essentially all observed at rest-frame 
angles $\sim 55-90^{\circ}$.  

In this context, it is interesting that the linear polarization structures
of 1055+018 and 1334$-$127 both correspond to the pitch-angle regime
that is expected to give rise to the observed transverse CP structure. 
1055+018 displays a ``spine+sheath'' linear polarization structure, with 
predominantly orthogonal inferred {\bf B} field along the jet ridge line 
and longitudinal {\bf B} field at the jet edges (Attridge et al.  1999, 
Pushkarev et al. 2005). This type of polarization structure will arise 
naturally if the jet has a helical {\bf B} field, if the pitch 
angle is relatively high and the viewing angle is not too far from 
$90^{\circ}$ in the jet rest frame (Lyutikov, Pariev \& Gabuzda 2005); 
the more symmetrical the CP structure,
the closer to a viewing angle of $90^{\circ}$ in the jet rest frame.
This is precisely the combination of pitch angle and viewing angle that
is required in the helical-field CP model to obtain roughly symmetrical
regions of positive and negative CP on opposite sides of the jet. 
1334$-$127 is one of a minority of AGN classified as BL Lac objects 
whose inner-jet {\bf B} fields are predominantly longitudinal,
indicating that, if its jet has a helical field, the pitch angle is not
greater than about $45^{\circ}$ (i.e., the longitudinal component of the
helical field is stronger than the toroidal component). Furthermore, 
the field is predominantly longitudinal in the VLBI core: 
quasi-simultaneously optical and 15+22+43-GHz VLBA polarization
observations showed that the optical and Faraday-rotation-corrected VLBI
core polarization angles were aligned to within $5^{\circ}$, with both
being orthgonal to the jet direction (Gabuzda et al. 2006); since the optical 
emission is clearly optically thin, this implies a longitudinal core 
{\bf B} field.  Thus, the observed transverse CP structure in 1334$-$127 
is also consistent with its linear polarisation structure: the simple
helical-field CP model of Gabuzda et al. (2008) predicts that CP of a 
single sign that is displaced toward one edge of the jet should be 
observed for jets whose helical fields have pitch angles somewhat less 
than $\simeq 45^{\circ}$ and are viewed at angles not too different from 
$\simeq 90^{\circ}$ in the jet rest frame. 

Thus, our CP images for 1334$-$127 and 1055+018 provide tantalizing 
evidence that a self-consistent picture of the observed intensity, linear 
polarisation and circular polarisation structures may be beginning to
emerge. While this is somewhat speculative at the moment, these results
certainly indicate the value of further high-resolution CP measurements,
as well as the development of further techniques for testing the reality
of tentatively detected transverse structures.

\section{Conclusion}

We have presented the results of parsec-scale circular polarization
measurements for 41 AGN at 15+22+43~GHz and an additional
18 at 15~GHz alone. We have detected parsec-scale CP in 8 sources for
the first time, and confirm previous detections in an additional 9 objects.
In all 7 AGN in which 15-GHz CP was detected in both our measurements and 
in the MOJAVE first-epoch measurements (Homan \& Lister 2006), the signs of 
the CP agree, demonstrating that a consistent picture is beginning to emerge 
from the collected results. 

The first-epoch MOJAVE measurements revealed the presence of jet CP in
five AGN (3C84, 3C273, 2128$-$123, 2134+004 and 2251+158). 
Vitrishchak \& Gabuzda (2007) reported the presence of jet CP in 1334$-$127 
and 3C279, and we have confirmed jet CP in these objects and added 
2223$-$052 to this list. 

These are among the very first multi-frequency VLBI CP measurements. We
have measured CP at more than one frequency for 10 of the 41 AGN.
No simple picture of the frequency dependence of the CP emerges.
For example, comparable numbers of the AGN displaying measureable CP
at both 15 and 22~GHz display higher $|m_c|$ values at the higher or
lower of these frequencies. At the same time, there is a clear tendency 
for the degrees of CP to be higher at 43~GHz than at the two lower 
frequencies. While virtually all the signs for the CP measured at 15 
and 22~GHz agree, the sign of the observed core-region CP often changes 
between 15/22 and 43~GHz (in 4 of 6 sources). This suggests the
action of optical-depth effects, or some other mechanism giving rise to
a systematic change in the CP sign with distance from the jet base. 

None of the observed CP spectra correspond in a straightforward
way to the frequency dependence expected for either intrinsic CP or
CP generated by Faraday conversion operating in a homogeneous source, 
which both predict that the CP should decrease with increasing frequency. 
The fact that we have found $m_c$ to increase with frequency in several 
sources, and to universally be higher at 43~GHz compared to our lower 
frequencies, is intriguing in 
light of the scaling arguments presented by Wardle \& Homan (2003), 
which point out that CP from either synchrotron radiation or Faraday 
conversion in a Blandford--K\"onigl jet could have an inverted CP spectrum, 
$m_c\sim\nu$. This suggests that our measurements are probing scales on
which the intrinsic inhomogeneity of the
jet plays an appreciable role in determining the observed CP and its
spectrum.  In addition, the picture may be complicated by the fact that
the core CP measurements may represent the superposition of CP contributions 
from several different regions in the core and innermost jet, as has been 
directly observed in the nearby radio galaxy 3C84 (Homan \& Wardle 2004).

Although we cannot distinguish between various CP-generation mechanisms
based on the observed spectra, we argue, as have previous authors,
that the synchrotron mechanism has trouble explaining the growing 
number of measurements indicating degrees of CP of the order of 1\%
or more. Therefore, it remains most likely that the detected CP is 
generated primarily by Faraday conversion (possibly with a smaller
contribution by the synchrotron mechanism in some objects). 
Gabuzda et al. (2008) have argued that the CP is generated
predominantly by Faraday conversion in helical {\bf B} fields inherently
associated with the jets.

We have found tentative evidence for transverse structure in the CP 
distributions of 1055+018 and 1334$-$127.  
In fact, transverse CP structures of the sort observed in 1334$-$127 and 
tentatively detected in 1055+018 would come about naturally if the CP is 
generated in helical jet {\bf B} fields. Further, this is true for both 
intrinsic CP and CP generated by helical-field-driven Faraday conversion. 
In contrast, CP generated in jets in which 
the unidirectional {\bf B} field giving rise to the CP is longitudinal 
would display transverse structure only by chance, due, for example, to 
bending of the jets. We thus consider the transverse CP structures 
tentatively revealed
by these first 43-GHz CP measurements to provide further supporting
evidence for the hypothesis that the CP of compact radio-loud AGN is
closely tied to helical {\bf B} fields that are organically related to
the jets.  

The interesting and unexpected new results we have presented here indicate
that further CP analyses at 22 and 43~GHz are very worthwhile, despite the 
techical difficulties they present
(weaker source fluxes, higher noise levels, particularly at 43~GHz). 
Further high-resolution
observations for additional AGN will indicate how common transverse CP
structures really are, and, in the case of symmetrical
transverse CP structures with opposite signs on opposite sides of the core 
region, more conclusively demonstrate their reality. One possibility in this
regard will be if similar structures can be detected further from the core,
as the reality of these can be tested directly using the PHC procedure 
described above and by Homan \& Lister (2006). 

\section{Acknowledgements}

We thank T. V. Cawthorne for useful discussions of this work, as well as
the anonymous referee for comments that were clear, very helpful and
submitted in a timely fashion. 






\end{document}